\documentclass[10pt,aps,prb,showpacs,superscriptaddress,twocolumn,citeautoscript]{revtex4-1}

\usepackage{graphicx,color}  
\usepackage{subcaption}
\usepackage{bm}        
\usepackage{amssymb,url}
\usepackage{amsmath}
\usepackage{hyperref} 
\usepackage{cases}

\def\eeq{\relax}
\def\beq#1#2\eeq{\begin{equation}\label{#1}#2\end{equation}}
\def\bal#1#2\eal{\begin{align}\label{#1}#2\end{align}}
\def\bse#1#2\ese{\begin{subequations}\label{#1}#2\end{subequations}}
\def\ba{\begin{aligned}}   \def\ea{\end{aligned}}

\def\dd{\operatorname{d}} 
\newcommand{\ii}{\ensuremath{\mathrm{i}}}

\def\Re{\operatorname{Re}} 
\def\Im{\operatorname{Im}}

\DeclareMathOperator{\spn}{span}

\begin{document}

\title{Metaclusters for the Full Control of Mechanical Waves}
\author{Pawel Packo}
\affiliation{Department of Robotics and Mechatronics, AGH - University of Science and Technology,
Al. A. Mickiewicza 30, 30-059 Krakow, Poland}
\author{Andrew N. Norris}
\affiliation{Mechanical and Aerospace Engineering, Rutgers University, Piscataway, NJ 08854-8058 (USA)}
\author{Dani Torrent}
\email{dtorrent@uji.es}
\affiliation{GROC, UJI, Institut de Noves Tecnologies de la Imatge (INIT), Universitat Jaume I, 12071, Castell\'o, (Spain)}
\date{\today}

\begin{abstract}
We present a new method for the control of waves based on inverse multiple scattering theory. Conceived as a generalization of the concept of metagrating, we call metaclusters to a finite set of scatterers whose position and properties are obtained by inverse design once we have defined their response to some external incident field. {The particular focus  is on designing passive metaclusters that do not require an external source of energy.} The method is applied to the propagation of flexural waves in thin plates, and to the design of far field patterns, although its generalization to acoustic or electromagnetic waves is straightforward. Numerical examples are presented to the design of uni and bidirectional ``anomalous scatterers'', which will bend the scattering energy along a specific direction, ``odd pole'' scatterers, whose radiation pattern presents an odd number of poles  and to the generation of vortical patterns. Finally, some considerations about the optimal design of these metaclusters are discussed.
\end{abstract}

\maketitle

\section{Introduction}\label{sec1}
Active and passive control of the energy transfer in electromagnetic and mechanical waves is a challenging problem with a large number of applications, such as focusing, imaging, beam forming, cloaking and energy harvesting, among others\cite{jin2019gradient}. The advent of  so-called metamaterials\cite{engheta2006metamaterials,deymier2013acoustic} provided a new perspective  since these artificial structures allow the design of materials with extraordinary properties capable of manipulating the flow of energy in ways that would be impossible with common materials, enlarging in this manner the number of devices for the control of electromagnetic and mechanical waves. 

More recently, the concept of ``metasurface", conceived as artificial planar metamaterials, has attracted an increasing interest. Being thinner and less dissipative than bulk metamaterials, these structures allow for more efficient ways of manipulating the wave energy, with the additional simplification in fabrication that planar structures present in comparison with bulk structures\cite{yu2011light,kildishev2013planar,yu2014flat}.

However, the major drawback of both metamaterials and metasurfaces is that their functionality is based on the extraordinary refractive/reflective properties they present, and most of the devices designed in this framework require  a large number of scattering elements in order to form an ``effective'' material whose effective physical properties provide metamaterials of their extraordinary properties. In the case of metasurfaces, the surface has to be gradually structured so that the effective gradient in the surface impedance allows for the manipulation of the energy flow. This large number of scattering elements is an important limitation in the efficiency of metamaterials and metasurfaces, since in practice the number of different scattering elements will be limited, especially in the micro or nano-scale.

To overcome these difficulties, several approaches have been explored recently to simplify the design of metasurfaces by means of diffraction gratings\cite{ra2017metagratings,wong2018perfect,torrent2018acoustic,packo2019inverse,popov2019constructing,popov2019designing}, in which it has been possible to find a complex scatterer or unit cell performing the same functionality as some metasurfaces.   However,  the design process is still complex and  functionality is limited to the control of the propagation direction of waves\cite{jin2019engineered,ni2019high,he2020experimental}. 

In this work, we present a generalization of the concept of a metagrating but for finite structures. The objective  is to show how, for a given incident field, we can obtain a cluster of scatterers and their physical properties such that the scattered field presents a pre-selected shape. If a particular diffraction pattern is desired for a specific type of incident wave,  we provide a method to design a cluster of scatterers capable of transferring the energy along the desired directions. The inverse design method presented is based on multiple scattering theory\cite{martin2006multiple} and the general principle is applicable to any kind of classical wave, including acoustic and electromagnetic waves. {We use flexural waves in plates as the model medium, due to its potential wide application, but the presented framework is general and applicable to wave scattering in other   media.} This work therefore provides  a general principle for the full control of mechanical and electromagnetic waves based on scattering elements. 

The paper is organized as follows: After this introduction, section \ref{sec2} develops the idea of the direct and inverse multiple scattering problem. Section \ref{sec3} explains how the method can be applied to the design of far field patterns and section \ref{sec4} shows numerical examples of specific patterns. Finally, section \ref{sec:summ} summarizes the work and some mathematical results are given in Appendices \ref{A} and \ref{B}.

\section{Direct and Inverse Multiple Scattering Problem}
\label{sec2}
When some incident field $\psi_0$ impinges on a cluster of $N$ point-like scatterers the total field $\psi(\bm r)$ can be expressed as the sum of the incident plus the scattered fields,
\beq{101}
\psi(\bm r)=\psi_0(\bm r)+\psi_s(\bm r). 
\eeq
 The scattered field is 
\beq{1012}
\psi_s(\bm r)=\sum_{\beta=1}^N B_\beta G(\bm r-\bm R_\beta),
\eeq
where 
 $G(\bm r) = G(|{\bf r}|)$ is the Green's function and the coefficients $B_\beta$ are obtained from the multiple scattering equation\cite{torrent2013elastic}
\beq{102}
\sum_{\beta=1}^N [t_\beta^{-1}\delta_{\alpha\beta}-G(\bm R_\alpha-\bm R_\beta)]B_\beta=\psi_0(\bm R_\alpha) . 
\eeq
This provides a system of $N$ equations with $N$ unknowns. The quantity $t_\alpha$ is the strength of each point-like scatterer and it is the only quantity that contains information about its physical properties.  
This describes the direct multiple scattering problem, in which the number of  scatterers, $N$,    their strengths $t_\alpha$ and locations $\bm R_\beta$ are known, from which  we  compute the $B_\alpha$  coefficients to finally determine the field in all of space. 

The inverse problem is as follows:  assume that the scattered field can be expressed as a linear combination of  basis functions $\phi_n$ such that
\beq{103}
\psi_s(\bm r)=\sum_{n=-\infty}^\infty  A_n\phi_n(\bm r),
\eeq
then we specify the inverse problem as determining a finite number $N_p$ of $A_n$ coefficients, for $n=1\ldots N_p$, so that the scattered field  will have a specified radiation pattern in the far-field.     In general there will be a matrix $S$ such that
\beq{104}
A_n=\sum_{\beta=1}^N S_{n\beta}B_\beta,
\eeq
therefore, if we select the number $N$ of particles in the cluster equal to the number $N_p$ of modes to design,  equation \eqref{104} constitutes a determinate system of $N$ equations with $N$ unknowns from which we can solve for the $B_\beta$ coefficients. Once these are known, we can obtain the $t_\alpha$  elements from equation \eqref{103} as
\beq{105}
t^{-1}_\alpha=\frac{1}{B_\alpha}\left(\psi_0(\bm R_\alpha)+\sum_{\beta=1}^NG(\bm R_\alpha-\bm R_\beta)B_\beta\right).
\eeq
Thus we can obtain the physical properties of each particle. The main challenge is to find a cluster configuration giving physically acceptable particles.

For the case of flexural waves  on thin elastic plates, $\psi$ is the plate deflection, $G$ is the solution for a point force per unit area applied in the positive $\psi$-direction, and  
\beq{25-}
B_\alpha = t_\alpha \psi( \bm R_\alpha),
\eeq 
is the point force per unit area of scatterer $\alpha$, see Appendix \ref{A}.   The parameter $t_\alpha$ is an effective  point impedance which can be interpreted in terms of a single degree of freedom system with mass, stiffness and damping. Physically acceptable particles cannot supply energy, i.e.\ they must be passive.   Assuming time dependence $e^{-\ii \omega t}$, the passivity constraints require that one or other of the following is met 
\begin{subnumcases}{}
\sum_{\alpha=1}^N \left( \Im t^{-1}_\alpha \right) |B_\alpha|^2 & $\leq 0$ ,\label{09}
\\
\Im t_\alpha^{-1} & $\leq 0$ .\label{09b}
\end{subnumcases}
 Equation \eqref{09} requires that the cluster be globally passive, meaning that some of the scatterers can provide energy but there should be a negative energy balance adding all the contributions of the scatterers. Equation \eqref{09b}, 
or equivalently $\Im t_\alpha \geq 0$,  is a more restrictive condition, since it requires that all scatterers be passive systems (see Appendix \ref{A} for details). The equality holds for zero  dissipation in both equations. The goal of the inverse multiple scattering problem is to obtain a set of particles all simultaneously satisfying the first or both constraints. For the first, global passivity, constraint we assume that although some scatterers may require energy supply, this energy can be transferred from other, locally passive, ones (see Appendix \ref{A}).

The specific problem addressed below is to engineer the cluster of point scatterers to provide a close approximation to a desired far field scattering response.  In the next section we outline the steps necessary to achieve this in an optimal sense.

\section{Far Field Engineering}\label{sec3}

\subsection{Direct far field solution}

The functions $\phi_n$ of \eqref{103} are chosen as the infinite set
\beq{-32}
\phi_n({\bf r}) =  G(\bm r) e^{\ii n \theta }, \ \ n\in \mathbb{Z},
\eeq
where the position is expressed in polar coordinates  ${\bf r} = (r,\theta)$ with respect to an origin at $r=0$.  This allows to uniquely identify the coefficients  $A_n$ of \eqref{103} as   far field amplitudes of the scattered wave.  
In order to see this, first note that the far-field  for a source at ${\bf R}_\beta = (R_\beta,\theta_\beta)$ is 
\beq{-1}
G(\bm r-\bm R_\beta) \approx G(\bm r) e^{-\ii k R_\beta\cos(\theta - \theta_\beta)} . 
\eeq
This  approximation holds whether the Green's function is  for the Helmholtz equation or for the Kirchhoff plate equation. In both cases, the far-field response depends only on the large argument approximation of $H_0^{(1)}(x)$. The scattered far-field of the cluster follows from \eqref{1012} and \eqref{-1} as 
\beq{-21}
\psi_s (\bm r)
\approx  G(\bm r) f(\theta) ,
\eeq
with the far-field radiation function 
\beq{-22}
f(\theta) = \sum_{\beta =1}^N B_\beta e^{-\ii k R_\beta\cos(\theta - \theta_\beta)} . 
\eeq
Alternatively, 
\beq{-25}
f(\theta) = \sum_{n=-\infty}^\infty A_n e^{\ii n \theta } ,
\eeq
where the infinite set of coefficients $\{A_n \}$ is related to the $N$ coefficients $\{B_\beta \}$
by \eqref{104} with  
\beq{-4}
S_{n\beta} = (-\ii)^n e^{-\ii n\theta_\beta}  J_n(kR_\beta) . 
\eeq

For a unit amplitude incident plane wave propagating in the direction $\theta = 0$ the radiation pattern function satisfies the optical theorem \cite{norris1995scattering}
\beq{7-3}
\Im f(0) = \sigma_\text{sca} +\sigma_\text{abs}   ,
\eeq
where the scattering cross-section $\sigma_\text{sca}$ and absorption cross-section $\sigma_\text{abs}$ are defined in Eq.\ \eqref{3=3}.  Further details can be found in  Appendix \ref{A}.  The cross-sections can also be expressed directly in terms of the coefficients $\{A_n \}$ and $\{B_\beta \}$, see Eq.\ \eqref{3=4}, leading to the explicit form of the optical theorem
\beq{7-4}
\Im f(0) =  \frac 1{8 D k^2}  {{\bf A}}^\dagger {\bf A} + \sum_{\alpha =1}^N 
\big(-\Im t_\alpha^{-1} ) |B_\alpha |^2  .
\eeq

{We define the energy efficiency of a cluster as the ratio of scattered to total input energy, which can be calculated from the
scattering and absorption cross-sections of Eqs. \eqref{3=3} as 
\beq{ma7}
\eta = \frac{\sigma_{sca}}{\sigma_{ext}} =  \frac{\sigma_{sca}}{\sigma_{sca} + \sigma_{abs}}.
\eeq}

\subsection{Inverse problem}\label{secinv}

In the inverse source problem we are given $f(\theta)$ and seek the cluster that optimally   reproduces this scattering pattern.  The   radiation pattern, defined by the coefficients $\{A_n\}$ in the form \eqref{-25}, is infinite dimensional, whereas the cluster comprises a  finite set of $N$ sources.  We define the error
\bal{-5}
E &= \int_0^{2\pi} \left| \sum_{n=-\infty}^\infty  
\big( A_n -  \sum_{\beta =1}^N S_{n\beta}B_\beta\big) e^{\ii n\theta } \right|^2
\notag
\\ &= || {\bf A} - {\bf S}{\bf B}||^2,
\eal
where $||  {\bf X}||^2 =   {\bf X}^\dagger {\bf X}$ with ${\bf X}^\dagger$ the Hermitian transpose of vector ${\bf X}$. Minimizing $E$ for given ${\bf A} $ and $ {\bf S}$ yields the solution 
\beq{6}
{\bf B} = \big( {\bf S}^\dagger {\bf S}\big)^{-1}  {\bf S}^\dagger  {\bf A}, 
\eeq
where $\big( {\bf S}^\dagger {\bf S}\big)^{-1}  {\bf S}^\dagger $ may be identified as the Moore-Penrose inverse of ${\bf S}$. 

The approximated radiation pattern is $f^{(N)}(\theta) $ 
\beq{-7}
f^{(N)}(\theta) = \sum_{n=-\infty}^\infty A_n^{(N)}  e^{\ii n \theta } , 
\eeq
where $A_n^{(N)}$, $n\in \mathbb{Z}$, are the elements of 
\bal{8}
{\bf A}^{(N)} &= {\bf S}{\bf B}
\notag \\
&= {\bf P} {\bf A}  ,
\eal
and  the    non-negative definite Hermitian matrix $\bf P$ is
\beq{8.1}
{\bf P} ={\bf S} \big( {\bf S}^\dagger {\bf S}\big)^{-1}  {\bf S}^\dagger .
\eeq
It is shown in Appendix \ref{B} that the  matrix  ${\bf P} $ is infinite dimensional    but  finite 
rank  with $N$ non-zero eigenvalues equal to $+1$, see Eq.\ \eqref{4=54}.    
It therefore acts as a projection from  the infinite dimensional space of far-field pattern functions to the $N-$dimensional set of approximate pattern functions: $f(\theta) \to f^{(N)}(\theta)$.

The optimal solution \eqref{8} yields an error  
\beq{9}
E =   {\bf A}^\dagger \big( {\bf A} - {\bf A}^{(N)}\big)
= ||  {\bf A}||^2  -|| {\bf Q} {\bf A}||^2  ,
\eeq
where  
\beq{12}
 {\bf Q} = \big( {\bf S}^\dagger {\bf S}\big)^{-\frac 12}  {\bf S}^\dagger . 
\eeq 
In practice we will be interested in the relative error
$E_\text{rel} = E/||  {\bf A}||^2$, i.e. 
\beq{87}
E_\text{rel} =   1  -\frac{ {\bf A}^\dagger   {\bf A}^{(N)} }{||  {\bf A}||^2} . 
\eeq

\subsubsection{Invisibility?}

{Can the cluster be invisible, in the sense that there is no scattered wave?   Setting 
${\bf A}$ to zero implies 
\beq{-34}
0 = {\bf S}{\bf B} 
\ \ \Rightarrow \ \ {\bf S}^\dagger {\bf S}{\bf B} =0. 
\eeq
Hence $B_\alpha =0$, and therefore $t_\alpha =0$, meaning there are no scatterers, the null solution. 
We conclude that the inverse scattering cluster scheme does not provide a useful route to invisibility or cloaking. 
}

\subsection{Inverse design algorithm}\label{invdes}

Based on the above findings, the inverse scattering design can be formulated as follows. 

\begin{enumerate}

\item The $N$ scatterer positions ${\bf R}_\alpha$, $\alpha = 1, \ldots , N$ are defined. 

\item The desired far field pattern $f(\theta)$ is specified, or equivalently the set of far field modal amplitudes $\{ A_n,\ n\in \mathbb Z\}$ are given (see \eqref{-25}).  

\item Frequency (equivalently wavenumber $k$) is given.

\item The matrices $\bf S$     and $\bf P$ are evaluated (see \eqref{-4}, \eqref{8.1}). 

\item The source strengths $B_\alpha$, the optimal approximation to the far field pattern $f^{(N)}(\theta)$, i.e.\ $\{ A_n^{(N)},\ n\in \mathbb Z\}$, and the relative error $E_\text{rel} $ are calculated (see \eqref{-25}, \eqref{6}, \eqref{87}).

\item An incident wave field $\psi_0({\bf r})$ is defined, and the particle impedances 
$t_\alpha$, $\alpha = 1, \ldots , N$, are 
calculated (see \eqref{105}). 

\end {enumerate}

The first two items are geometrical, independent of frequency and the incident wave.  Once the frequency is defined, the 
approximation $f^{(N)}(\theta)$ to the scattered far field is optimal in the sense of an $N$-dimensional solution according to the setup, and it is independent of the incident field.  The form of the incident wave, combined with the source amplitudes $B_\alpha$,   defines the required particle impedances $t_\alpha$ in Eq.\ \eqref{105}.  

{The inverse algorithm defines the  configuration mechanical properties, i.e. the $t_\alpha$, for a given incident wave $\psi_0$.  If the incident wave changes, then the new scattering coefficients $B_\alpha$ are defined by the system of equations \eqref{102} with the predetermined $\{t_\alpha \}$.  Regardless of the incident wave direction, the process remains {\em reciprocal} under the interchange of incident and scattering directions. 
}

The question that must be addressed is whether or not all of the scatterer impedances  satisfy the passivity constraints \eqref{09} or \eqref{09b}.

\section{Applications}\label{sec4}

\subsection{Far field patterns and the matrix $\bf P$} \label{kkay}

Two groups of cluster patterns are considered, namely regular polygons, where scatterers are uniformly distributed over a circle, and finite lattices, where scatterers are regularly distributed in a 2D finite grid.   We describe how different arrangements of the scatterers influence the matrix $\bf P$ of Eq.\ \eqref{8.1} which defines the optimal approximation to the desired scattering pattern.

\subsubsection{Scatterers on a regular polygon}

Let us assume that the $N$ scatterers lie on the circle of radius $R$ at $\theta_\beta = 2\pi \beta/N$.  We consider $\bf A$ corresponding to each of the modes $e^{\ii m \theta}$, $m\in \mathbb Z$, so that 
$||  {\bf A}||^2 = 1$ and $E \le 1$ with $E \ll 1$ indicating the desired scattering mode is well approximated.  The results of numerical experimentation are as follows. For small $kR$ relative to $N$,  $E$ is small for modes $m=0 , \pm 1, \ldots , \frac{N-1}2$ if $N>1$ is odd, and for modes $m=0 , \pm 1, \ldots , \frac{N-2}2$ if $N$ is even, with 
$E\approx 0.5$ for $m=\pm \frac N2$.  The accuracy diminishes as $kR $ increases.  In other words, for small $kR$ the $N$ unit eigenvalues of $\bf P$ correspond to modes $m=0 , \pm 1, \ldots , \frac{N-1}2$ if $N>1$ is odd, with analogous association for $N$ even.  Since the modes are multiply degenerate (all of eigenvalue unity) it follows that any linear combination of  these modes is an eigenvector. 

\subsubsection{Scatterers on a finite square lattice}

We now assume  that the scatterers are distributed in a square but finite lattice. The lattice is $M\times M \equiv N$ with lattice spacing $a$.  For instance, with $M=3$ and $ka=1$ we find that the 9 eigenvectors of $\bf P$ of  eigenvalue $+1$ span the space 
$\Omega_4 \equiv \{ 
e^{\ii m \theta}, \ m=0, \pm 1 , \ldots , \pm 4 \}$.  This result is arrived at by inspecting the error $E$ for each mode, and noting that it is small, on the order of $1e-4$ typically, while higher modes have error of approximately unity.  

However, for the same $ka=1$ but the larger lattice with $M=4$ $(N=16)$ we find that the nontrivial eigenspace is $\Omega_5  \bigcup \Omega_{6,10}$ where $\Omega_{6,10}$
is a five dimensional subspace formed from $\{ 
e^{\ii m \theta}, \ m= \pm 6 , \ldots , \pm 10 \}$.  

\subsubsection{General properties of the ${\bf P}$ matrix}
Numerical experiments on matrix ${\bf P}$ for different spatial configurations of the clusters show that for large and moderate $k R$ there are exactly $N$ eigenvalues of ${\bf P}$ with values close to 1. For large $k R$, the corresponding eigenvectors (i.e.\ patterns of scattering modes of the cluster) are highly irregular and sensitive to both $k R$ and scatterers positions (while the number of eigenvalues of value $1$ equals $N$). For $k R_c \approx 0.5$ and smaller, where $R_c$ is the characteristic size of the cluster, the eigenvalues of ${\bf P}$ begin to differ and assume values other than 1. For $k R_c \ll 1$ the number of nonzero eigenvalues reduces and low order scattering patterns are preferred.

Some general remarks on the number of scatterers ($N$) and scattering properties of the cluster can be formulated as follows. The larger $N$, the larger number of cluster modes, thus more complex scattering patterns can be reproduced accurately. Large number of scatterers in the cluster, on the other hand, may result in overconstraining the minimization problem and lack of locally or globally passive solutions. For moderate $k R_c$ typically the number of regular patterns (eigenvectors of ${\bf P}$) is similar to $N$, while more degenerate patterns and/or smaller number of similar eigenvalues ($\approx 1$) are observed for large or small $k R_c$.

\subsection{Scattering patterns}\label{scattpatt}

The inverse design of metaclusters is illustrated with the scatterers arranged on regular polygons or square lattices, as outlined in Sec. \ref{kkay}. Here we present the target scattering patterns that will be later reproduced by proper selection of passive impedances.

\subsubsection{Uni- and bi-directional scattering patterns}
\begin{figure}[h]
	\centering
	\subcaptionbox{Uni-directional at $3/4 \pi$ \label{wea} } 
	{\vspace{-.21cm}\includegraphics[width=0.49\linewidth ]{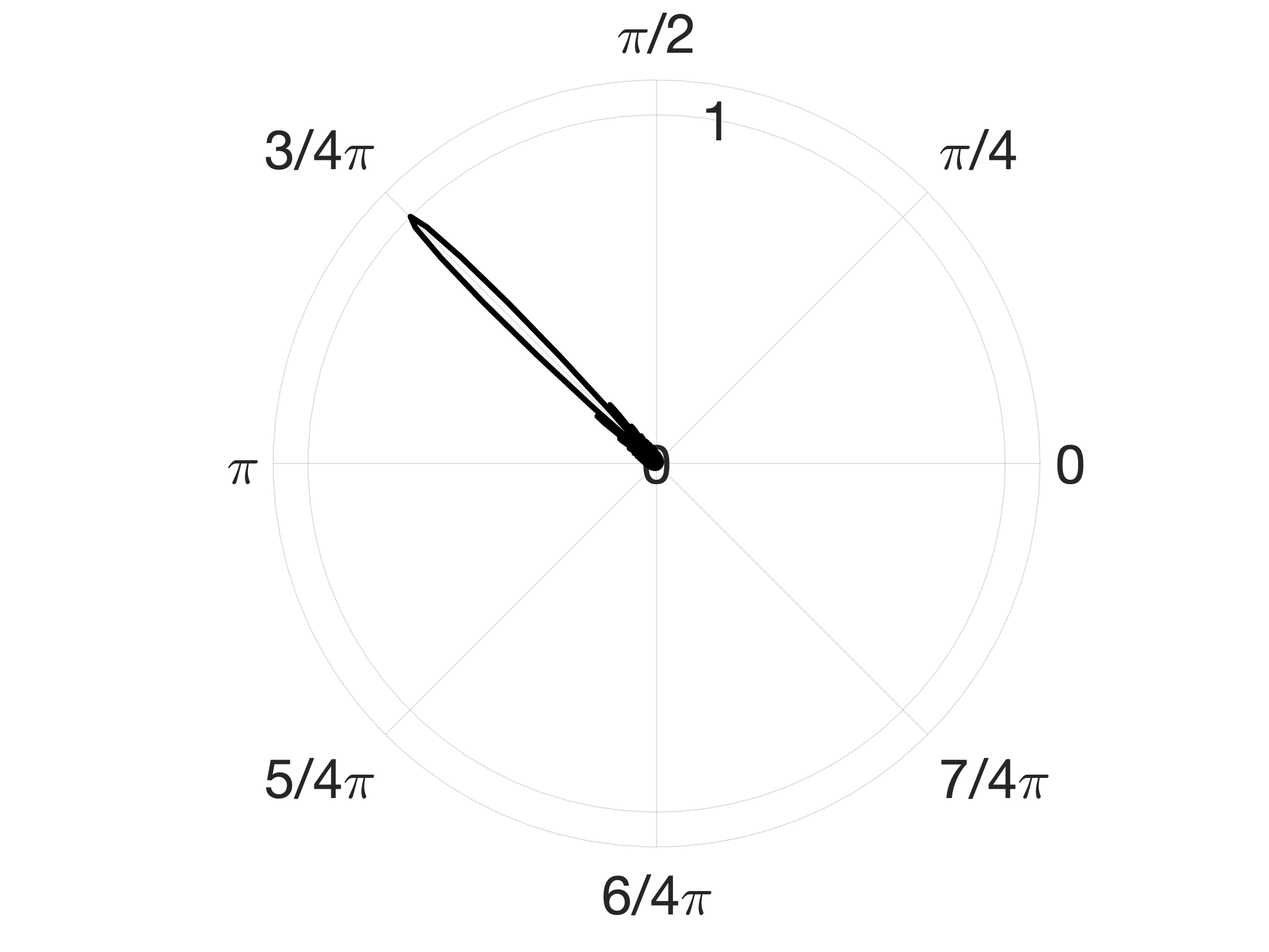}}
	\subcaptionbox{Bi-directional symmetric at $3/4 \pi$ \label{web} } 
	{\vspace{-.21cm}\includegraphics[width=0.49\linewidth  ]{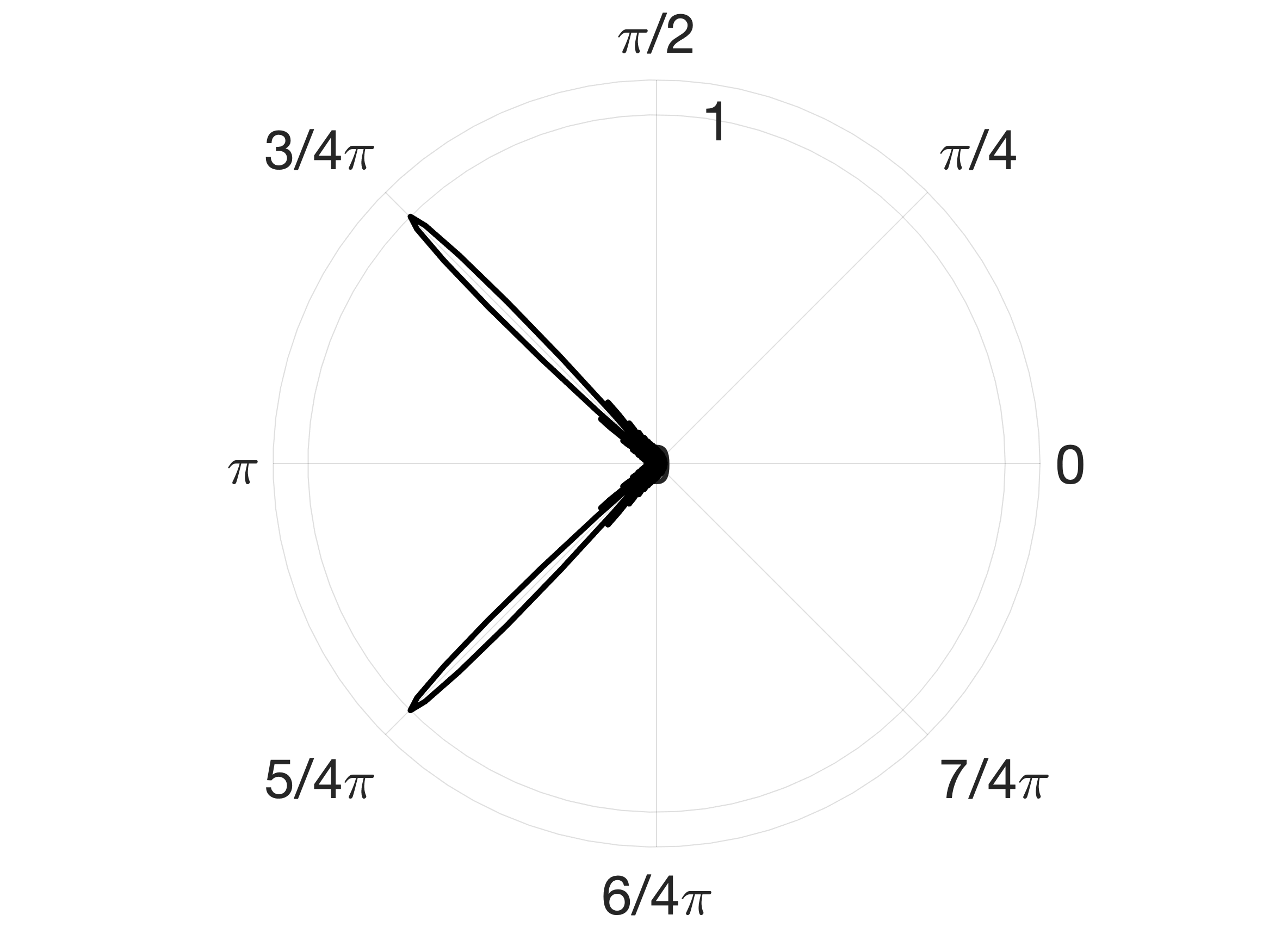}}
	\subcaptionbox{Tripole \label{wec}} 
	{\vspace{-.21cm}\includegraphics[width=0.49\linewidth] {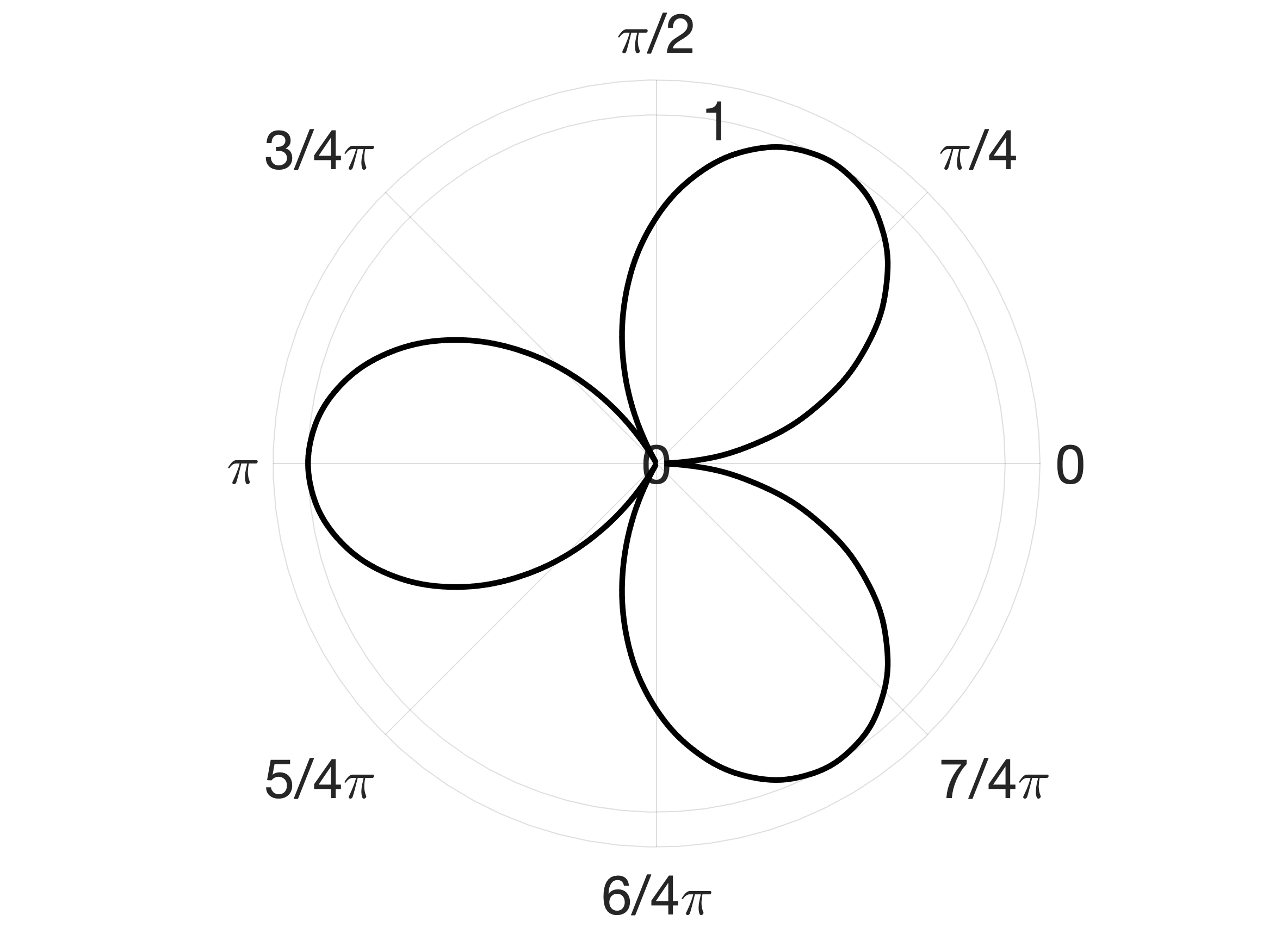}}
	\subcaptionbox{Pentapole  \label{wed} } 
	{\vspace{-.21cm}\includegraphics[width=0.49\linewidth]{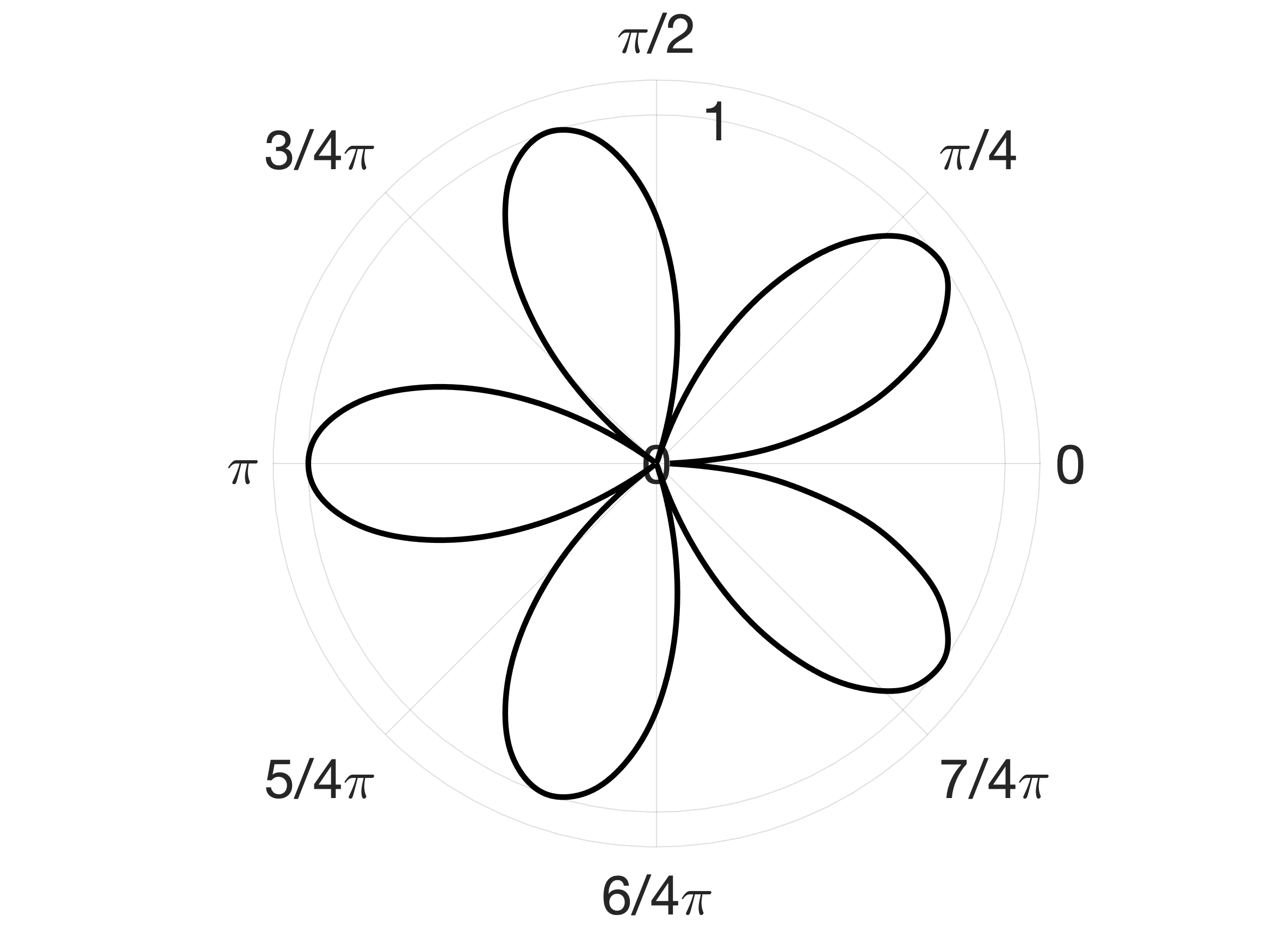}}
	\subcaptionbox{Vortex \label{wee} } 
	{\vspace{-.21cm}\includegraphics[width=0.49\linewidth]{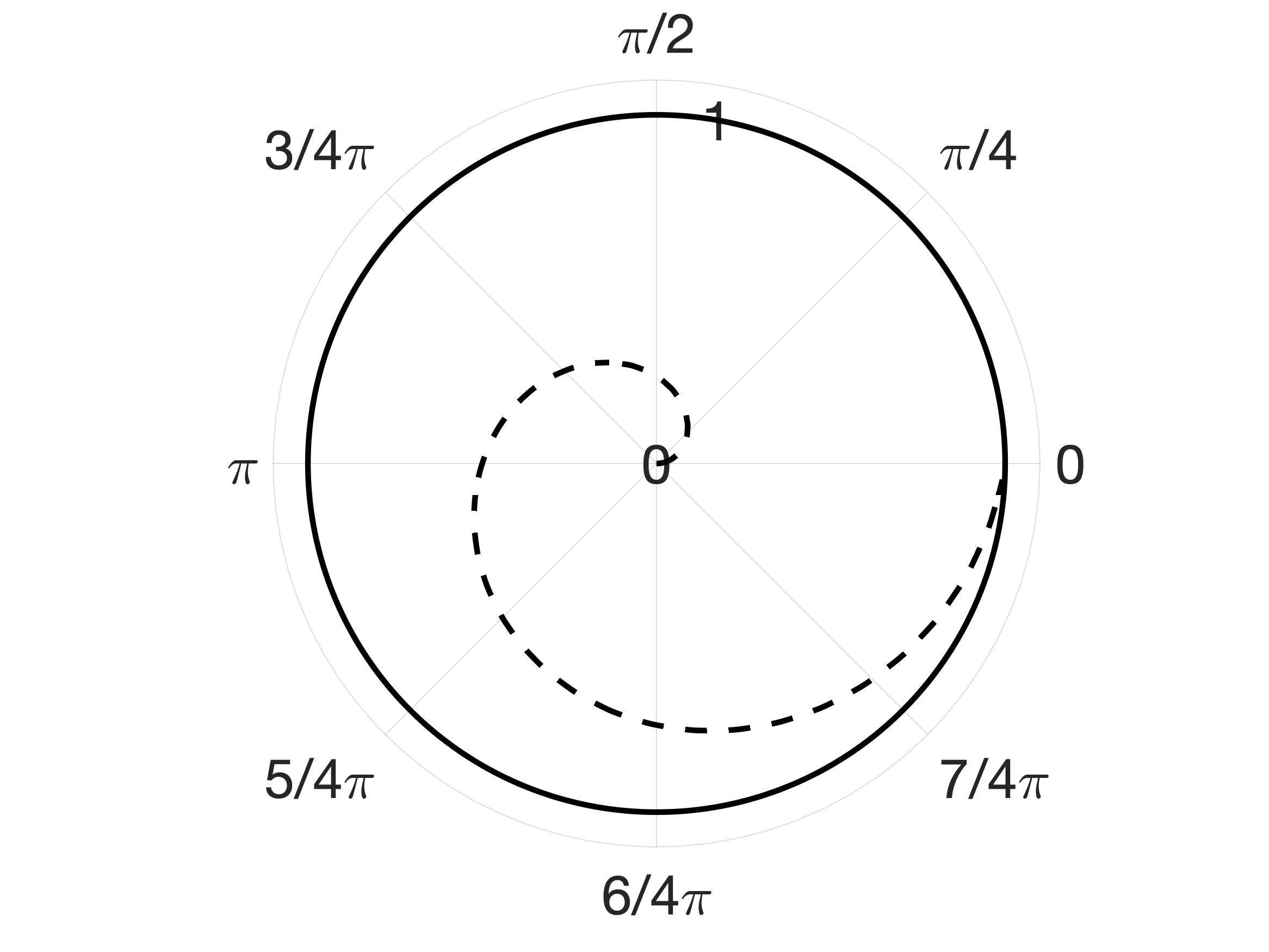}}
	\caption{Examples of the target patterns. Solid lines are normalized amplitudes, dashed lines are normalized phases of the pattern functions (finite number of $81$ modes in Eq.\ \eqref{-25} was assumed). }
	\label{fig1sdf}
\end{figure}

Uni-directional scattering in the direction $\theta = \theta_0$ corresponds to 
\beq{310}
f(\theta) = C_0 \delta(\theta  - \theta_0)
\  \ 
\Leftrightarrow \ \  
A_n = \frac {C_0}{2\pi} e^{-\ii n \theta_0}.
\eeq   A bi-directional scattering pattern is of the 
form $f(\theta) = C_0 \delta(\theta  - \theta_0) +  C_1 \delta(\theta  - \theta_1)$.  We consider patterns that are symmetric or anti-symmetric about the $x-$direction $(\theta = 0)$, 
corresponding to $\theta_1 = - \theta_0$ and $C_1 = \pm C_0$. We may choose $C_0 = 1$ with no loss in generality, and define 
\beq{31}
f_\pm (\theta , \theta_0)\equiv 
\delta(\theta  - \theta_0) \pm \delta(\theta  + \theta_0)
\ \ 
\Leftrightarrow \ \ 
A_n = \begin{cases}
\frac 1{\pi} \cos n \theta_0 ,
\\
\frac {-\ii}{\pi} \sin n \theta_0 .
\end{cases}
\eeq
Examples of the uni- and bi-directional scattering patterns are shown in Figs. \ref{wea} and \ref{web}. . 

\subsubsection{Odd-pole patterns}

Odd-pole patterns have $\bar{p}$ scattering lobes directing energy towards that preferential directions. The odd-pole scattering pattern and the corresponding $A_n$ coefficients are given as
\beq{590}
f(\theta) = \sin \left( \frac{\bar{p}}{2} \hspace{2pt} \theta '\right)
\  
\Leftrightarrow \ 
A_n = \frac{\bar{p} \sin^2 \big( \frac{\bar{p}}{2} +n\big) \frac{\pi}2     }{\pi \left( \left( \frac{\bar{p}}{2} \right)^2 - n^2 \right)}  
\eeq
{where $\theta ' = \theta \,\text{mod}\, 2\pi$ is used to ensure that  $f(\theta)$ is a $2\pi-$periodic function. }

\paragraph{A tripole}

For a tripolar pattern, $\bar{p} = 3$, there are three main lobes spaced every $2 / 3 \pi$. An example of a tripolar pattern is shown in Fig. \ref{wec}.

\paragraph{A pentapole} Similarly, for $\bar{p} = 5$, a pentapole scattering pattern is obtained. Figure \ref{wed} illustrates this type of pattern.

\subsubsection{A vortex}

A vortex generates uniform constant amplitude pattern with angle-dependent linearly changing phase behavior. The corresponding formulas for the vortex of order $\bar{p} \in \mathbb{Z}$ are

\beq{59x0}
f(\theta) = e^{\ii \bar{p} \theta}
\  
\Leftrightarrow \ 
A_n =   \delta_{ n\bar{p} }.
\eeq

Directional characteristics of amplitudes and phases for the vortex pattern are shown in Fig. \ref{wee}.

\subsection{Full metacluster designs}

Designing a metacluster requires finding all $t_\alpha$ for a given cluster topology and the desired scattering pattern. The procedure outlined in Sec. \ref{invdes} is employed here to find $t_\alpha$. We first present metacluster scattering patterns corresponding to the desired patterns from Sec. \ref{scattpatt}, obtained for different clusters configurations. Since the inverse procedure frequently leads to active particles, we next impose the condition \eqref{09b} to find locally passive optimal metaclusters and present their scattering responses.
For all presented examples we introduce the incident wave - without loss of generality - assumed to be a plane wave in the $-x$ direction $(\theta = \pi)$. 

\subsubsection{Scattering patterns for optimal metaclusters}

\begin{figure}[h]
	\centering
	\subcaptionbox{Uni-directional at $3/4\pi$ \label{1a} } 
	{\vspace{-.21cm}\includegraphics[width=0.49\linewidth ]{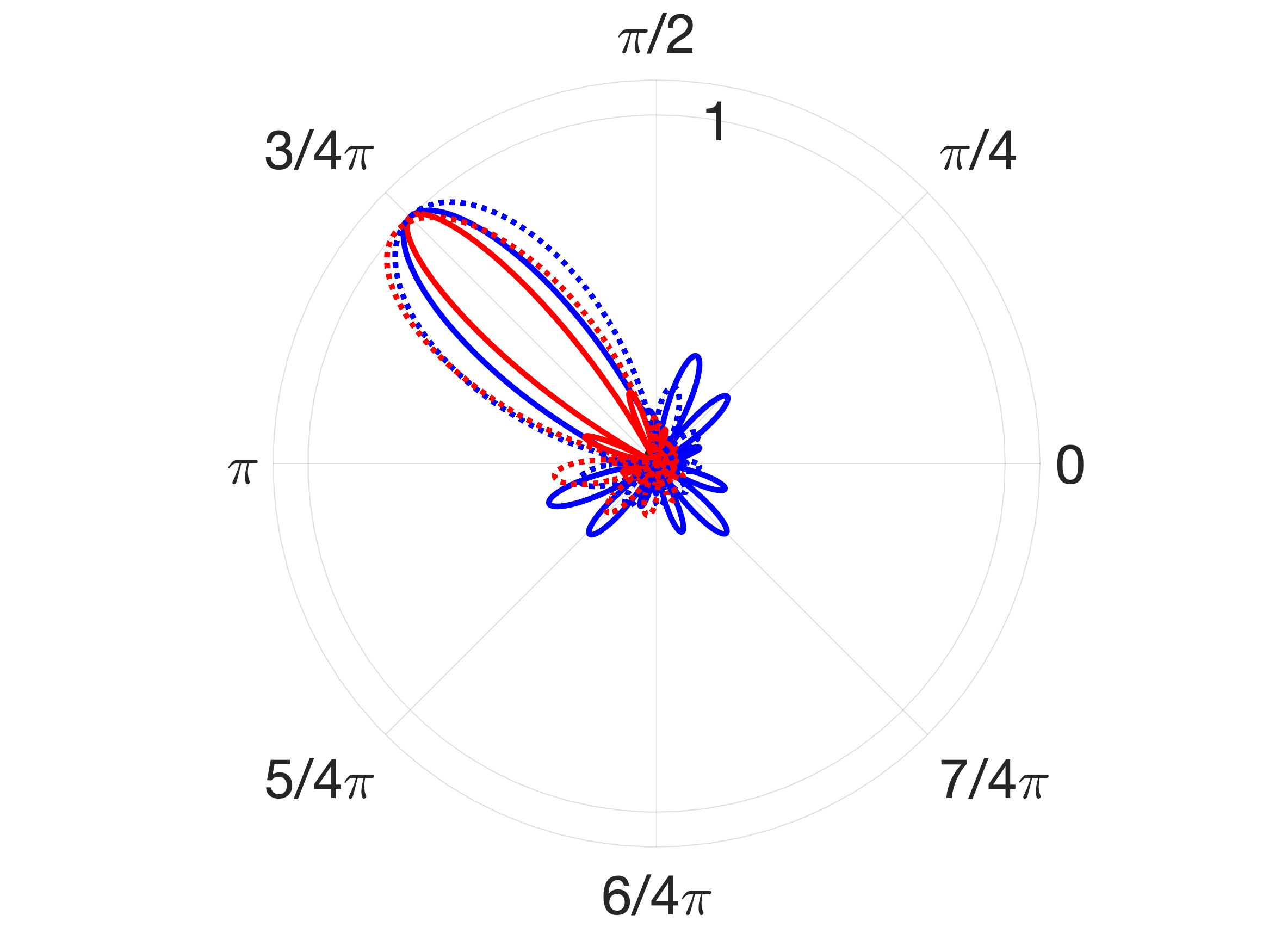}}
	\subcaptionbox{Bi-directional symmetric at $3/4\pi$ \label{1b} } 
	{\vspace{-.21cm}\includegraphics[width=0.49\linewidth  ]{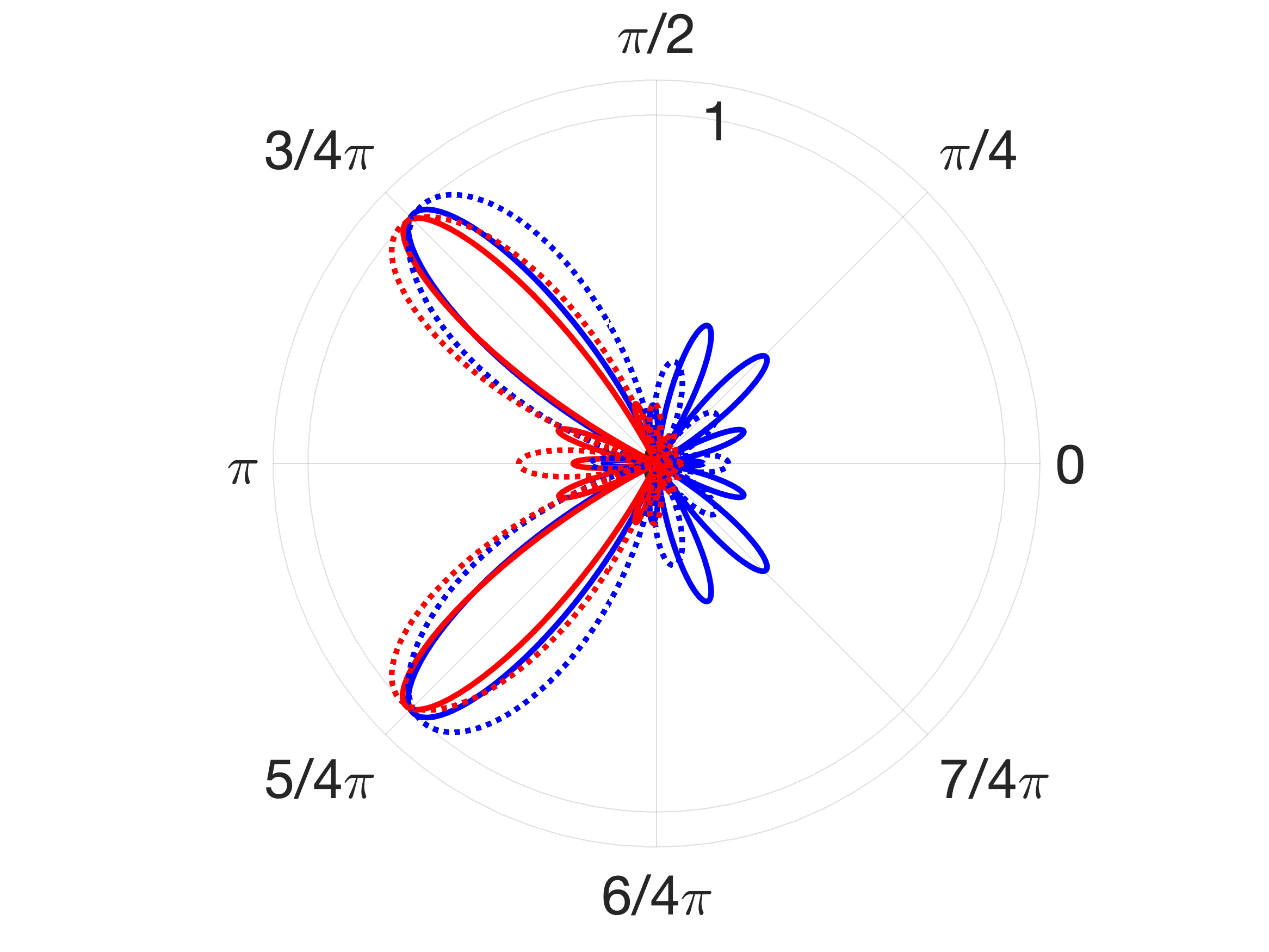}}
	\subcaptionbox{Tripole \label{1c}} 
	{\vspace{-.21cm}\includegraphics[width=0.49\linewidth] {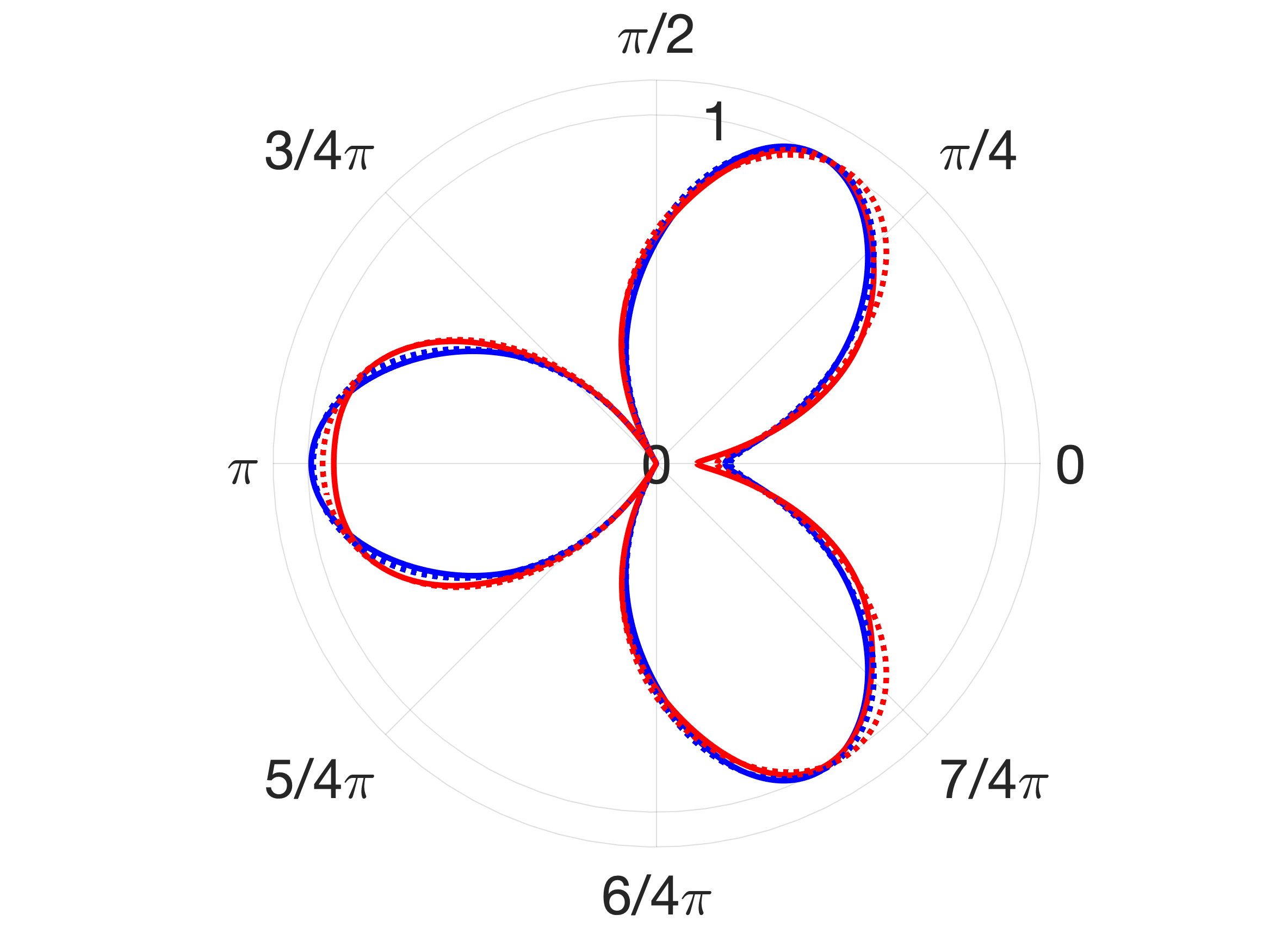}}
	\subcaptionbox{Pentapole  \label{1d} } 
	{\vspace{-.21cm}\includegraphics[width=0.49\linewidth]{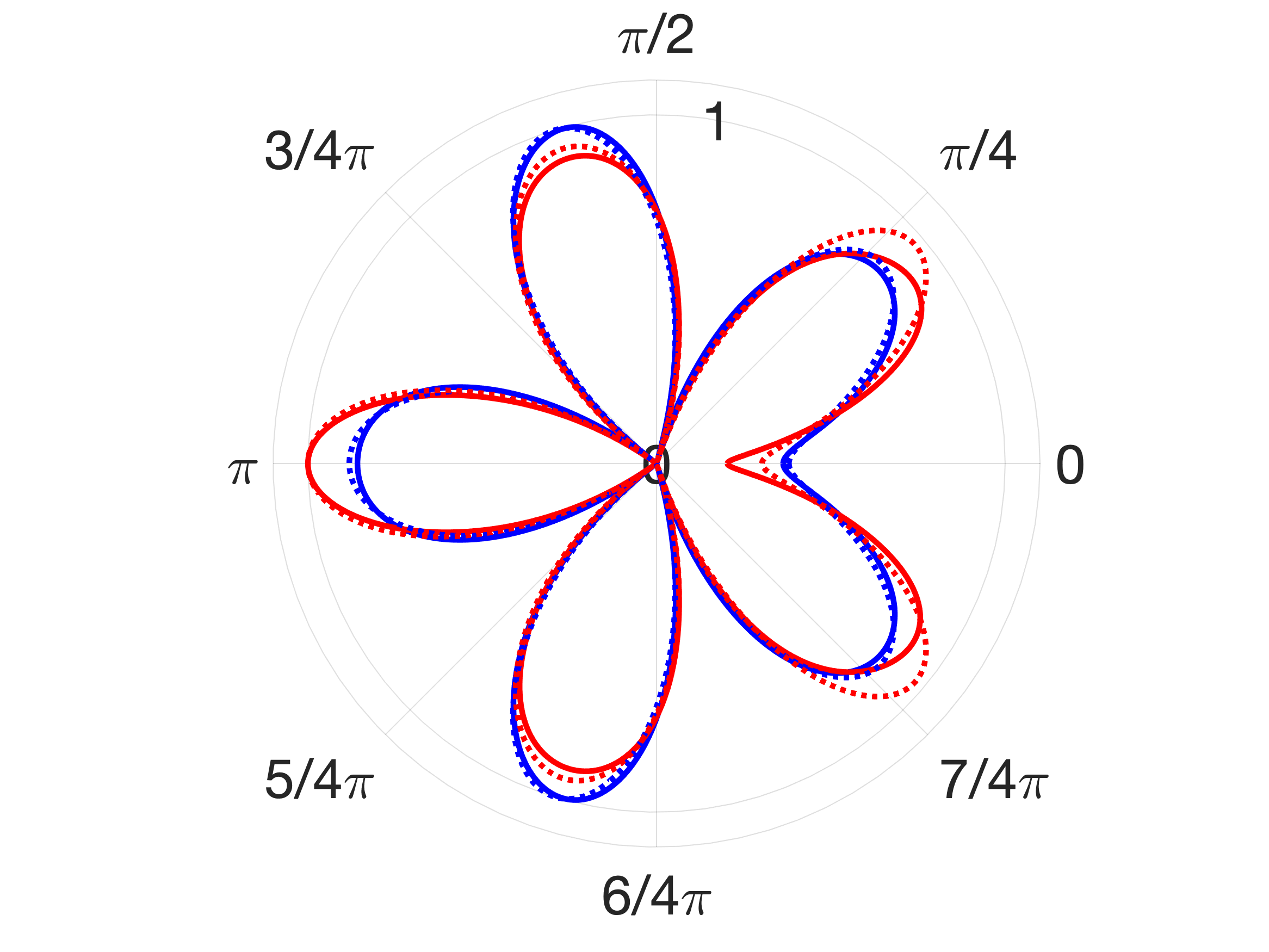}}
	\subcaptionbox{Vortex  \label{1e} } 
	{\vspace{-.21cm}\includegraphics[width=0.49\linewidth]{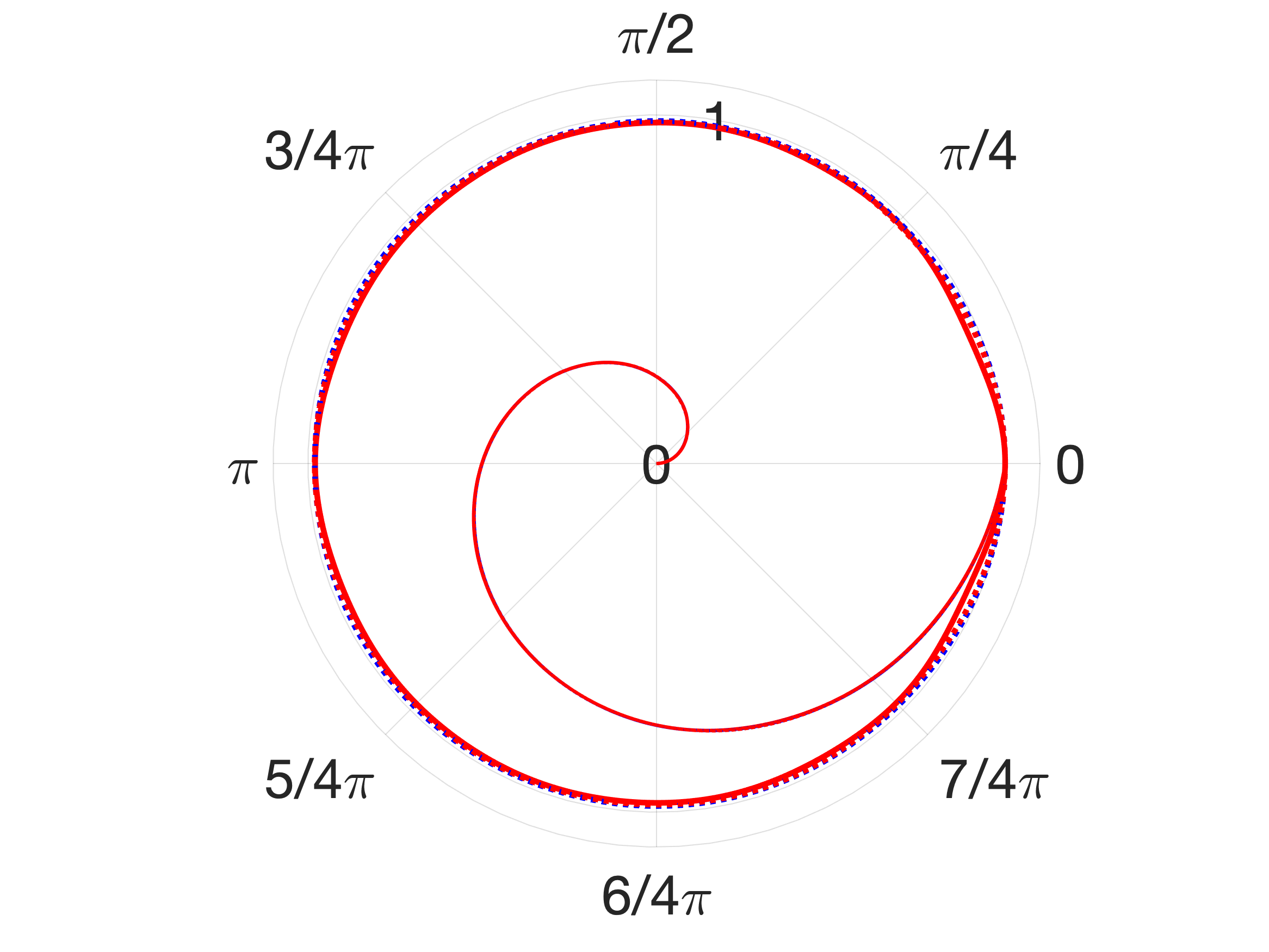}}
	\subcaptionbox{} 
	{\vspace{-.21cm}\includegraphics[width=0.20\linewidth]{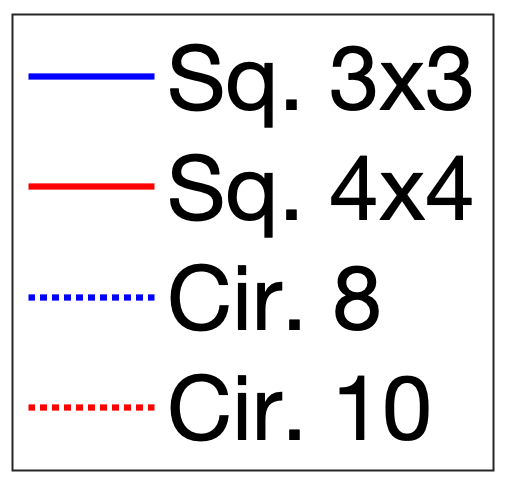}}
	\caption{Examples of the optimal far field patterns for square {$3\times3$} and {$4\times4$} arrays (lattice spacing $a$) and circular arrangements with $8$ and $10$ particles (radius $a$), for $ka = 1$ based on Eqs.\ \eqref{-7} and \eqref{8}.  Incidence angle $\theta = \pi$ (the $-x$ direction). For the vortex, Fig.\ \ref{1e}, circular shapes are amplitude profiles while the spirals in the center are phases of the scattering pattern. All patterns are normalized.}
	\label{fig1}
\end{figure}

Scattering patterns obtained for selected cluster topologies are shown in Fig. \ref{fig1}. Very good agreement between the desired patterns of Fig. \ref{fig1sdf} can be seen, proving the effectiveness of the design procedure. However, some of the corresponding impedances - computed using the inverse approach of Sec. \ref{invdes} - are active, hence require energy supply. We next analyze and adopt the inverse procedure for seeking only locally passive solutions.

\subsubsection{An optimization problem for passive metaclusters}

{Our design objective is the set of point impedances $\{ t_\alpha, \, \alpha = 1, \ldots , N\}$. We aim at fulfilling the local passivity condition, Eq.\ \eqref{09b}.
Define
\beq{-67}
\ba
u_\alpha &= t_\alpha^{-1}, 
\\
p_\alpha &= B_\alpha^{-1} \psi_0(\bm R_\alpha), 
\\
s_\alpha &= B_\alpha^{-1}\sum_{\beta=1}^NG(\bm R_\alpha-\bm R_\beta)B_\beta,
\ea
\eeq
then  Eq.\ \eqref{105} becomes
$
u_\alpha = p_\alpha +s_\alpha$, $\alpha = 1, \ldots , N $.
Consider plane wave incidence  
$
\psi_0({\bm r}) = p_0 e^{\ii {\bm k}\cdot {\bm r}}$, 
for some wavenumber $\bm k$. 
There is a further degree of freedom that has not been used.   This could be considered as the
  amplitude and phase of the incident wave, i.e.\ the complex number $p_0$.   Alternatively, if we fix $p_0 = 1$, then there is a similar degree of freedom in how we normalize the far field pattern function $f(\theta)$.  This has the effect of scaling $\bf A$ and hence $\bf B$ by a complex number.  This scaling redefines $p_\alpha$ but has no effect on $s_\alpha$ of \eqref{-67}. }
	
Therefore, with no loss in generality we assume the incident wave has unit amplitude, 
\beq{-69}
\psi_0({\bm r}) = e^{\ii {\bm k}\cdot {\bm r}}, 
\eeq
and rewrite  Eq.\ \eqref{105}, { the solution of the inverse problem, to incorporate this added degree of freedom,} as
\beq{-68}
u_\alpha = z p_\alpha +s_\alpha, \ \alpha = 1, \ldots , N .
\eeq
Here the complex number $z$ defines the scaling of the pattern function, which goes as $z^{-1}$.  The important point is that $z$ can be chosen arbitrarily; in particular, we use it as an optimization parameter.   The fact that the pattern function amplitude is inversely proportional to $|z|$ for unit amplitude incident wave suggests that smaller $|z|$ is preferred {for maximizing efficiency of energy conversion}. 

The optimization problem is as follows: given the $N$ complex numbers $p_\alpha$ associated with the incident wave and the $N$ complex numbers $s_\alpha$ associated with the point sources,
find $z$ {of Eq.\ \eqref{-68}} that ensures $\Im u_\alpha \le 0$ for all $\alpha$.  If this can be achieved then the optimal solution is the one with minimum value of $|z|$, ensuring maximum amplitude for the pattern function. 
It might not be possible using the single complex number $z$ to obtain all of 
the complex numbers $u_\alpha$ in the negative imaginary half-plane.  If this is not  achievable in practical examples then the constraint may be relaxed, for instance,  to minimize the maximum instance of positive $\Im u_\alpha$.  Then the "nearest" passive configuration can be identified by setting  $\Im u_\alpha$ to zero for those particles with positive $\Im u_\alpha$. Another alternative could be based on condition \eqref{09}, i.e.\ when the metacluster is globally passive - meaning that the net energy supplied to the cluster non-positive.

In cases where the search procedure for $t_\alpha$ failed to find locally passive metaclusters, a rigid rotation was applied to the cluster (equivalent to   changing the incidence angle) and the search was repeated.

\subsubsection{Example: A passive optimal metacluster for uni- and bi-directional patterns}

\begin{figure}[h]
	\centering
	\subcaptionbox{(normalized) Red: pattern at $ka=1.9$, blue: pattern at $ka=2.8$, black: target pattern \label{2a} } 
	{\vspace{-.21cm}\includegraphics[width=0.49\linewidth ]{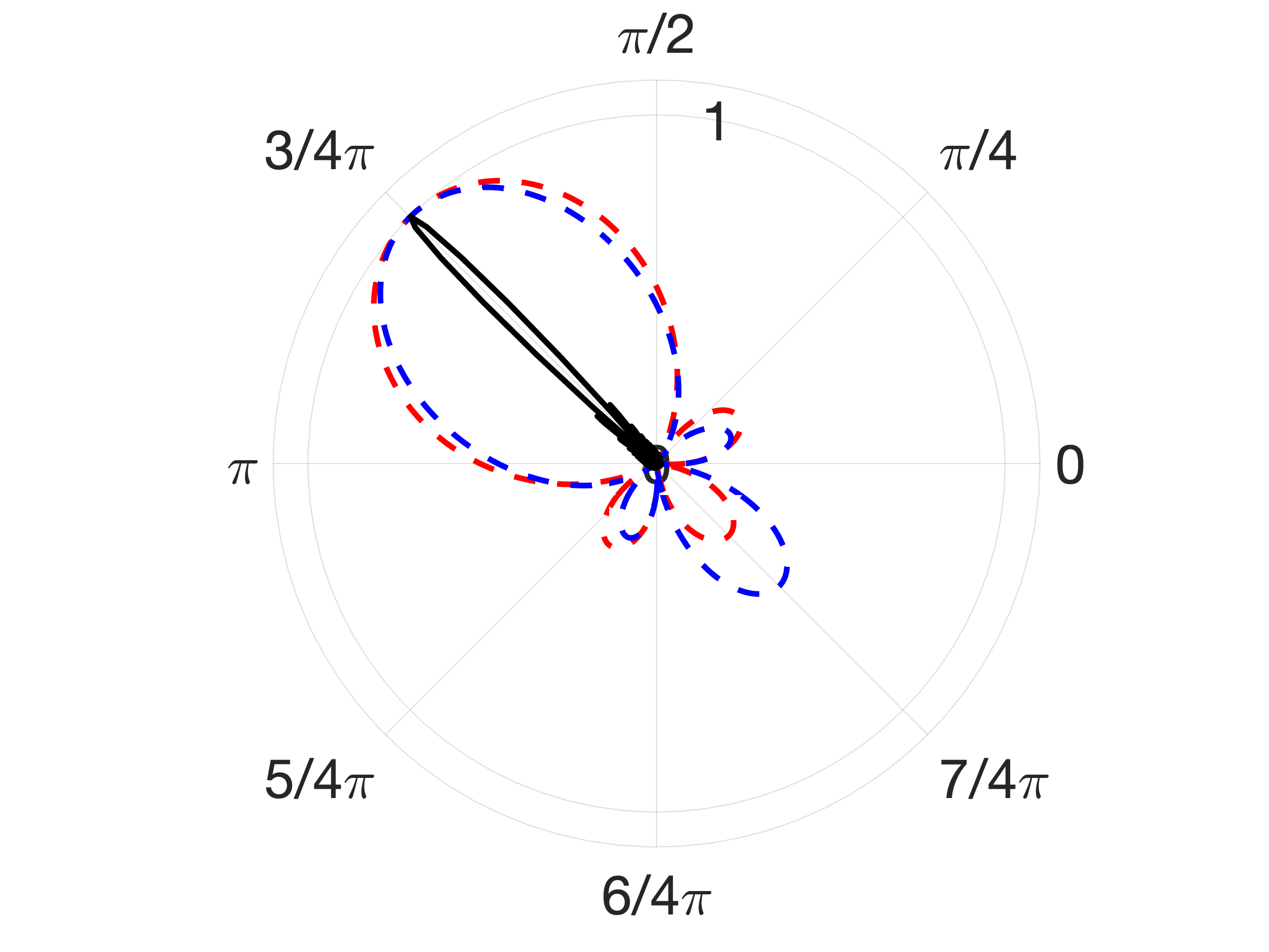}}
	\subcaptionbox{Displacement magnitude at  $ka=1.9$  } 
	{\vspace{-.21cm}\includegraphics[width=0.49\linewidth  ]{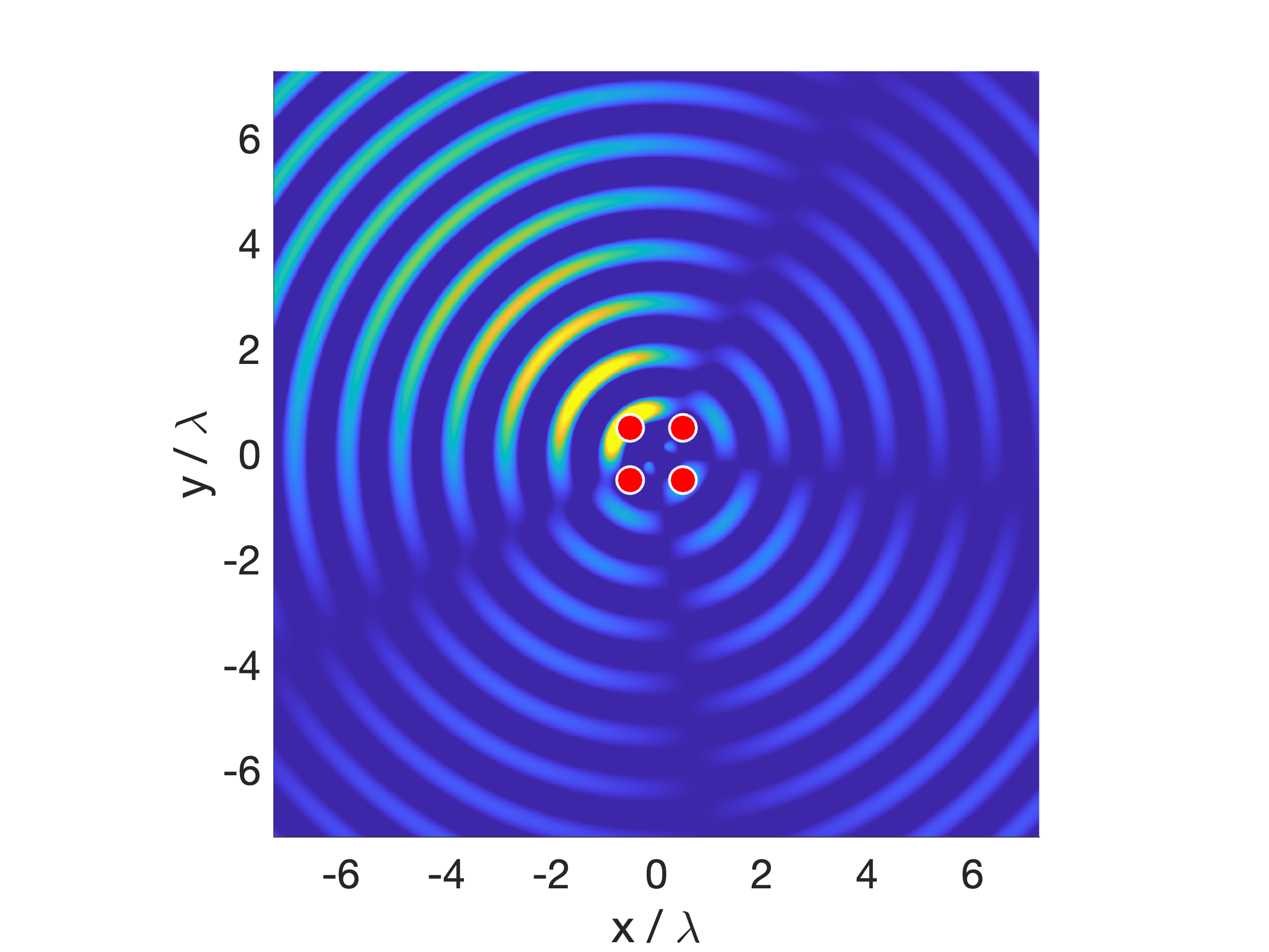}}
	\caption{The optimal pattern functions for the passive $2\times2$ metacluster at two frequencies bounding a bandwidth of passive designs, $ka \in \{ 1.9, 2.8 \}$, (a), and the corresponding displacement field generated at $ka = 1.9$, (b). Admittances of the cluster are give in Table \ref{table1}. {The efficiencies of energy conversion are $\eta_{ka = 1.9} = 0.60$ and $\eta_{ka = 2.8} = 0.35$.}}
	\label{fig2}
\end{figure}

Numerical experimentation shows there are metacluster configurations for which the inverse impedance solutions are all passive. Examples of the uni- and bi-directional scattering patterns for  a square lattice metacluster are shown in Figure \ref{fig1}. More detailed investigations show that - for instance - a square array with lattice parameter $a$ designed to direct a wave incident from the $\theta = \pi$ direction into a scattered wave preferentially directed toward $\theta = 3/4\pi$
has  totally passive solutions for $1.9 \le ka \le 2.8$.  The optimal passive admittances $t_\alpha^{-1}$ are frequency dependent, with values at the end of the passive interval shown in Table \ref{table1}.  The associated optimal scattering patterns are shown in Figure \ref{fig2}. In all examples we take $a=1$ and $D=1$.

The examples in Figure \ref{fig2} and Table \ref{table1} are based on the value of $z$ in \eqref{-68} for which the largest value of $\Im t_\alpha^{-1}$ is zero.  This optimizes the passive array in terms of its efficiency in converting the incident energy into a directed far field pattern.   The metacluster dissipates wave energy but in a way that is most efficient among all passive options. {For the cluster shown in Fig. \ref{fig2}, the values of the efficiency parameter  $\eta$ of Eq.\ \eqref{ma7} are $\eta_{ka = 1.9} = 0.60$ and $\eta_{ka = 2.8} = 0.35$.}

\begin{center}
\begin{table}
  \begin{tabular}{| c | c | c |}
    \hline
		 \hline
		${\bf R}_\alpha$  & $t_\alpha^{-1}$ & $t_\alpha^{-1} $ \\
		 & $ka = 1.9$ & $ka = 2.8$ \\
		 \hline
    $(-0.50,-0.50)$  &  0.0662 - 0.0050\ii  & 0.0438 - 0.0000\ii
   \\ \hline
    $(-0.50, 0.50)$  &   -0.0243 - 0.0000\ii & 0.0214 - 0.0618\ii 
		\\ \hline
    $( 0.50,-0.50)$  &    0.1120 - 0.0000\ii & -0.0651 - 0.1138\ii  
		\\ \hline
		$( 0.50, 0.50)$  &   -0.0413 - 0.0316\ii & -0.0524 - 0.0009\ii \\ 
    \hline
\end{tabular} 
\caption{The admittances  $t_\alpha^{-1}$ for the $2\times 2$ array of passive particles sending the wave incident at $\theta = \pi$ into the  $\theta = 3/4 \pi$ direction at two frequencies, see Figure \ref{fig2}.  } \label{table1}
\end{table}
\end{center}

\begin{figure}[h]
	\centering
	\subcaptionbox{(normalized) Red: pattern at $ka=3.1$, black: target pattern \label{24a} } 
	{\vspace{-.21cm}\includegraphics[width=0.49\linewidth ]{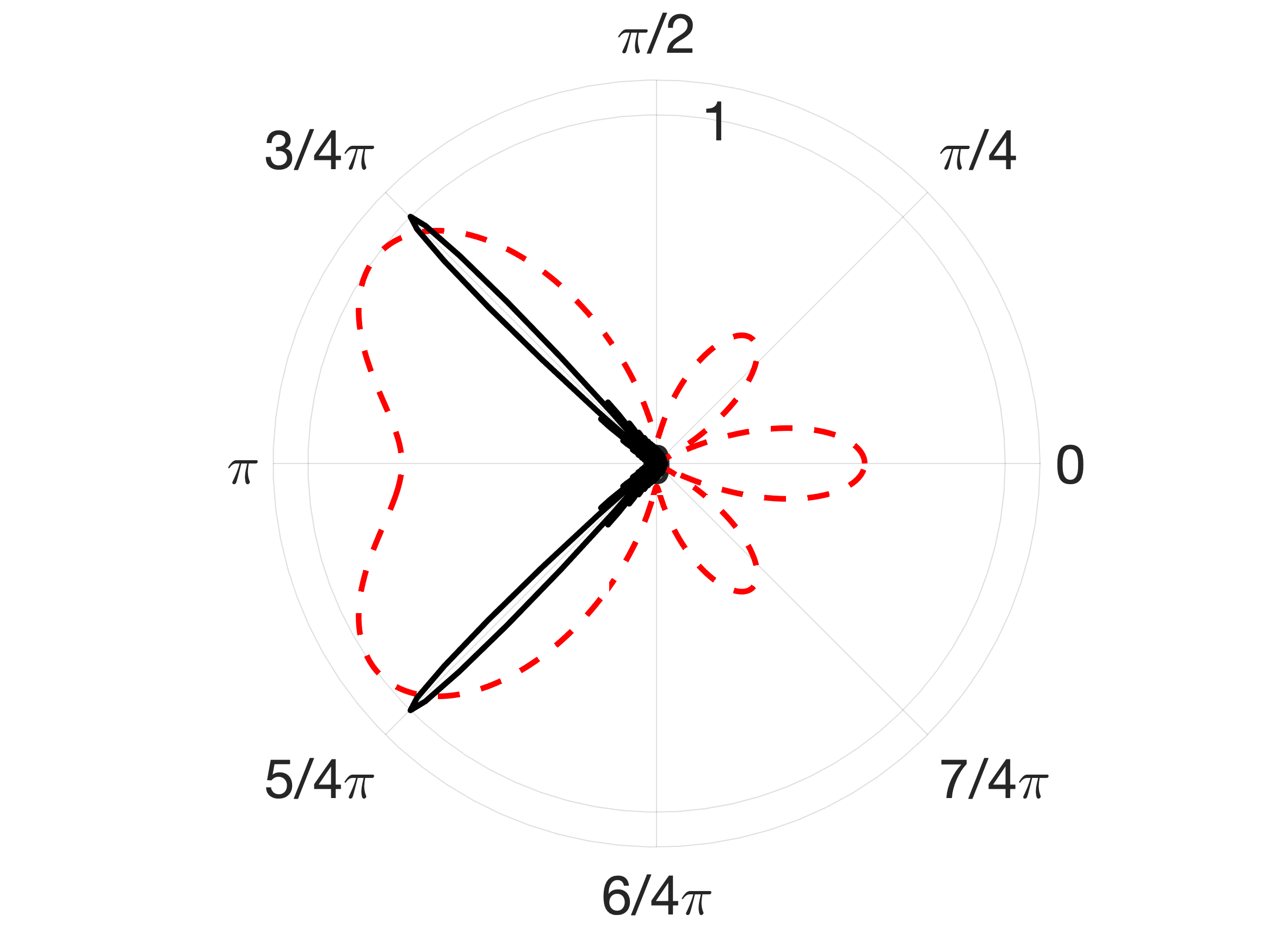}}
	\subcaptionbox{Displacement magnitude at $ka=3.1$  } 
	{\vspace{-.21cm}\includegraphics[width=0.49\linewidth  ]{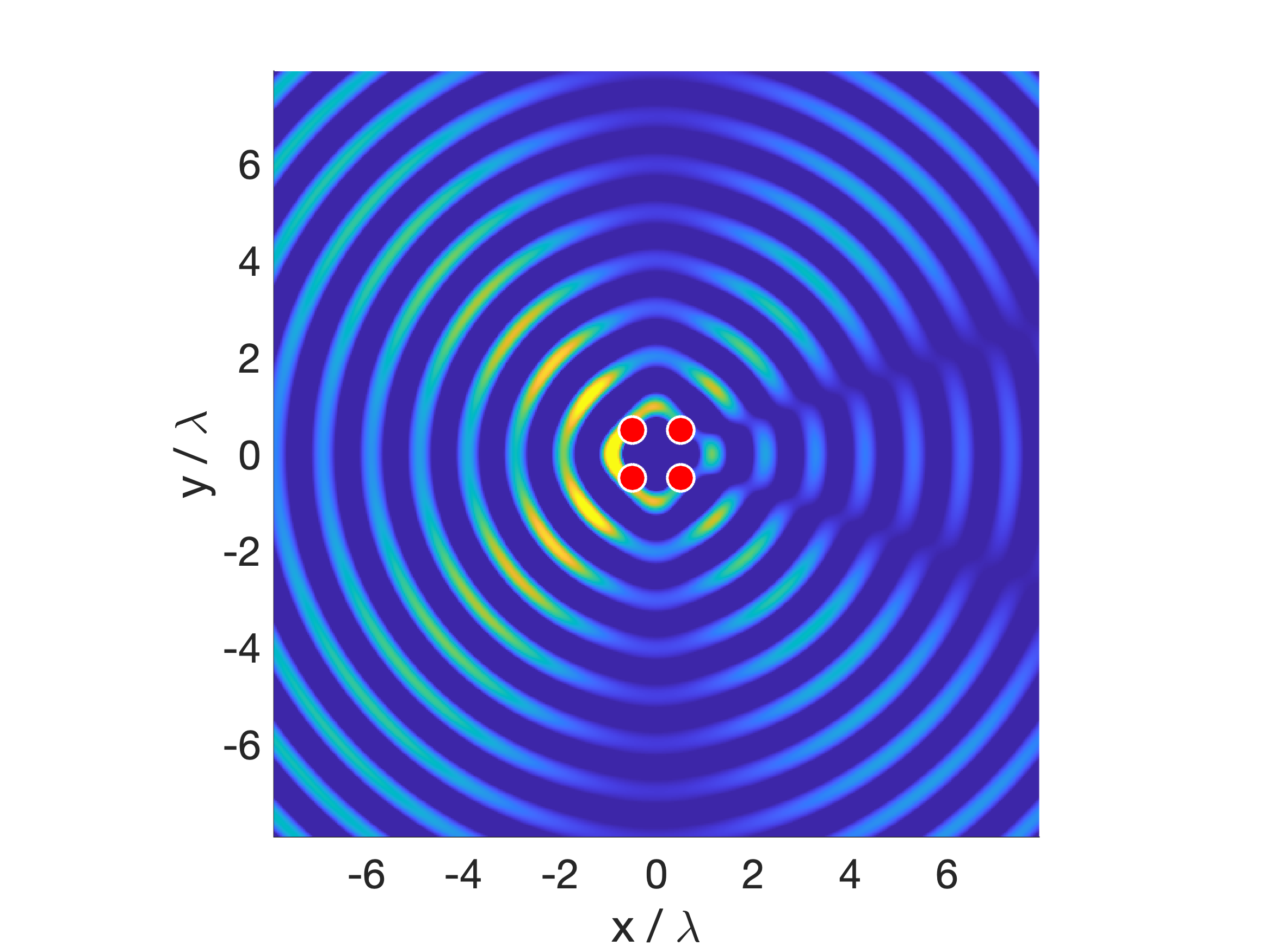}}
	\caption{The optimal pattern function for the passive $2\times2$  metacluster, (a), and the corresponding displacement field generated at $ka = 3.1$, (b). Admittances of the cluster are give in Table \ref{table14}. {Energy effiviency for this setup $\eta = 1$.}}
	\label{fig24}
\end{figure}

Similarly, Fig. \ref{fig24} shows a passive optimal $2\times2$ metacluster that converts the incident plane wave into a symmetric bi-directional pattern. Although the metacluster scattering pattern roughly approximates the desired far field function, all admittances are purely real indicating no dissipation in the system. {Consequently, the energy efficiency for this cluster is optimal, $\eta = 1$.}

\begin{center}
	\begin{table}
		\begin{tabular}{| c | c |}
			\hline
			\hline
			${\bf R}_\alpha$  & $t_\alpha^{-1}$ \\
			& $ka = 3.1$ \\
			\hline
			$(-0.50,-0.50)$  &  -0.0038
			\\ \hline
			$(-0.50,0.50)$  &   -0.0038
			\\ \hline
			$( 0.50,-0.50)$  &    -0.0111
			\\ \hline
			$( 0.50, 0.50)$  &   -0.0111 \\ 
			\hline
		\end{tabular} 
		\caption{The admittances  $t_\alpha^{-1}$ for the $2\times 2$ array of passive particles transforming the incident wave into a bi-directional symmetric pattern at $\theta = 3/4 \pi$, see Figure \ref{fig24}.  }   \label{table14}
	\end{table}
\end{center}

Further experimentation shows that obtained optimal solutions are very sensitive to the scatterers' positions and impedances. Also, requirements of symmetric clusters are overconstrained, most often resulting in at least one active particle, especially for large number of particles $N$.

\subsubsection{Example: A passive optimal metacluster for odd-polar patterns}

\begin{figure}[h]
	\centering
	\subcaptionbox{(normalized) Red: pattern at $ka = 5.4$, black: target pattern \label{33a} } 
	{\vspace{-.21cm}\includegraphics[width=0.49\linewidth ]{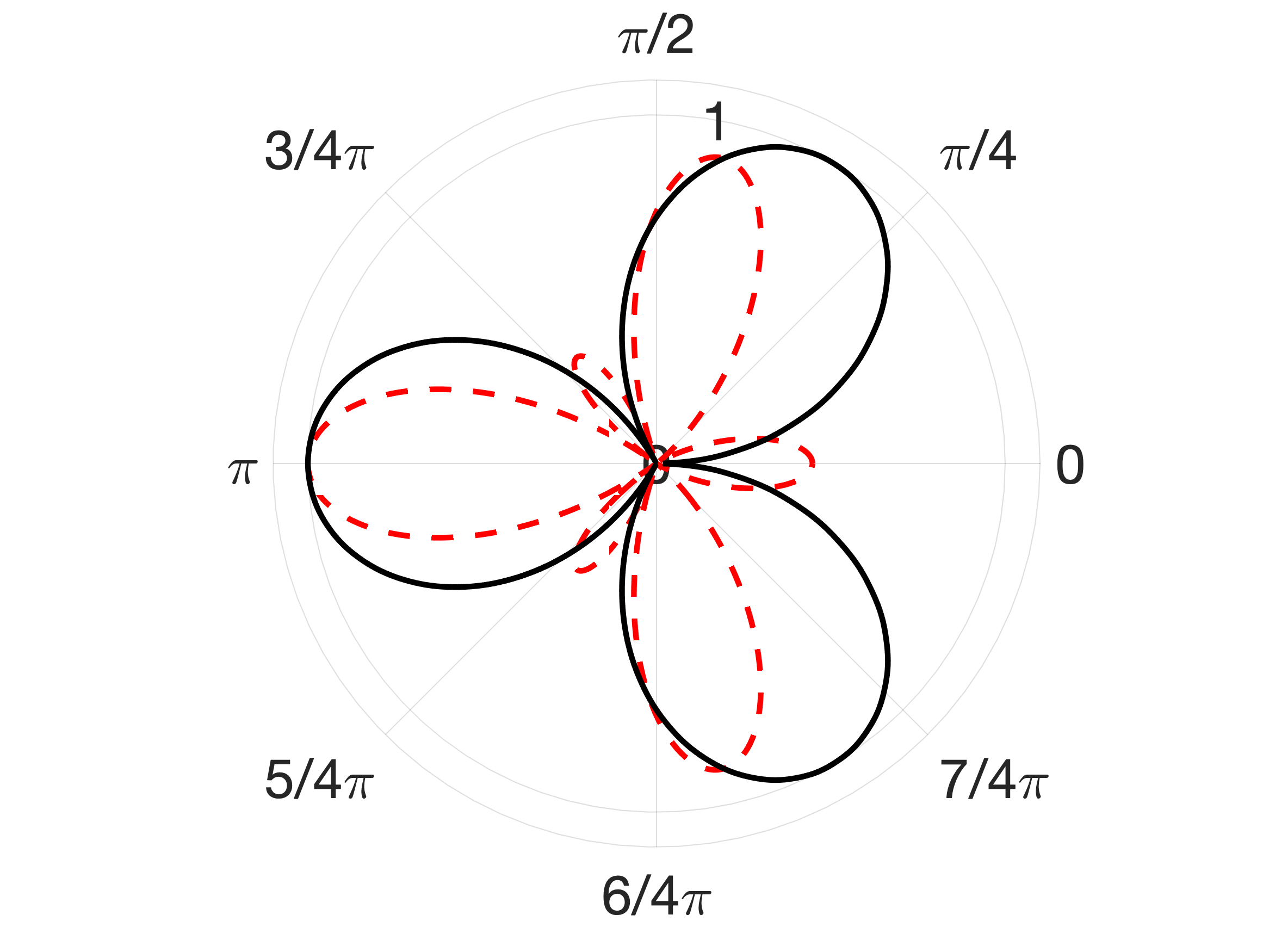}}
	\subcaptionbox{Displacement magnitude at $ka = 5.4$. \label{33b} } 
	{\vspace{-.21cm}\includegraphics[width=0.49\linewidth  ]{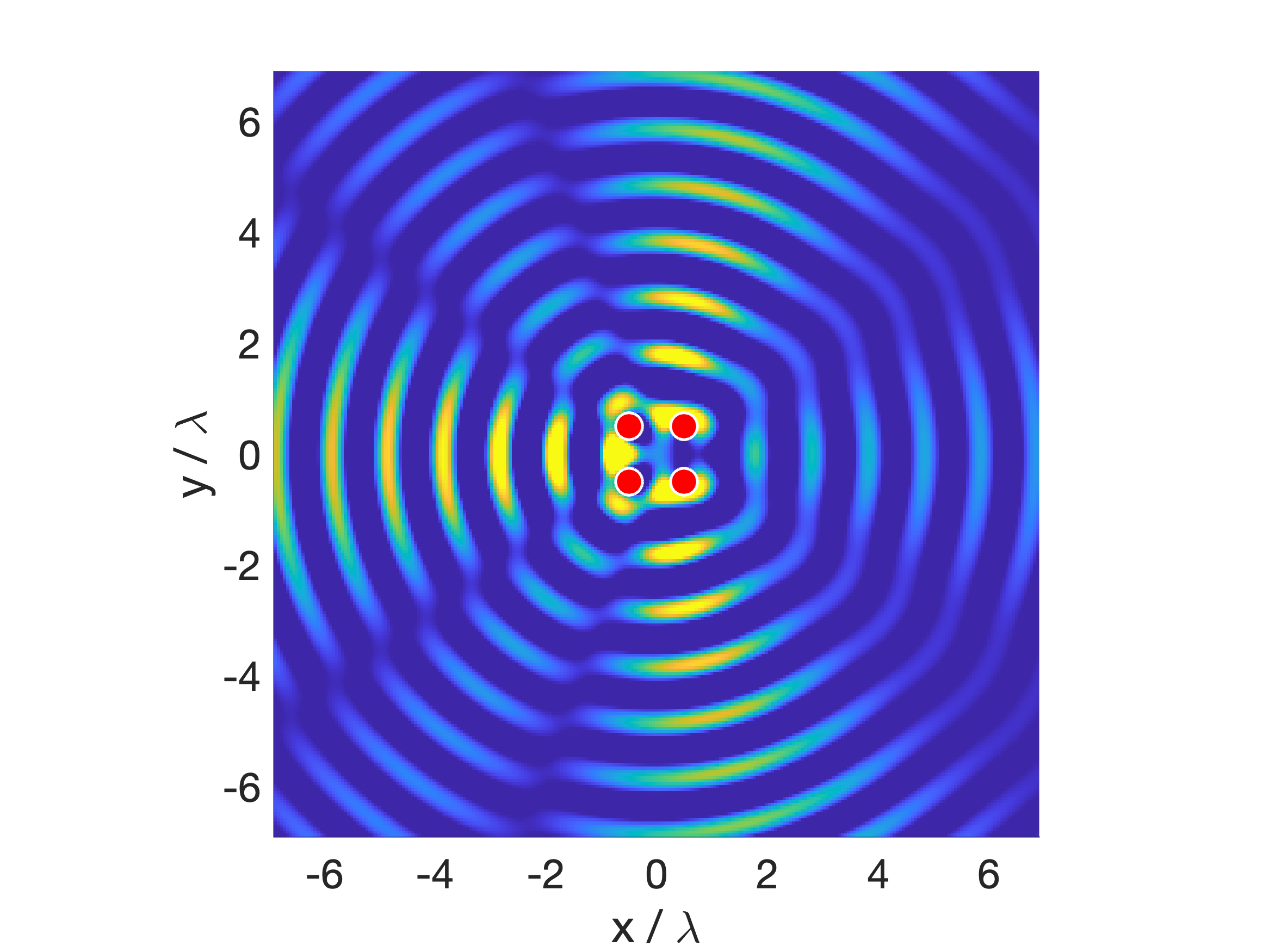}}
	\caption{The optimal pattern function for the passive $2\times2$  metacluster, (a), and the corresponding displacement field generated at $ka = 5.4$, (b). Admittances of the cluster are give in Table \ref{table2}. {The energy efficiency parameter is $\eta = 0.17$.}}
	\label{fig33}
\end{figure}

Analogously to the previous search, we look for optimal passive clusters capable of generating a scattering tripole. Figure \ref{fig33} shows the target and the actual scattering patterns for the tripole obtained for a square $2 \times 2$ cluster of scatterers. The optimal positions and admittances of the scatterers are shown in Table \ref{table2}. The admittances have nearly the same passive damping properties. The corresponding displacement field pattern generated by the metacluster is shown in Fig. \ref{fig33}. {The energy conversion efficiency is $\eta = 0.17$.}

\begin{center}
	\begin{table}
		\begin{tabular}{| c | c |}
			\hline
			\hline
			 ${\bf R}_\alpha$ & $t_\alpha^{-1} $ \\
			 & $ka = 5.4$ \\
			\hline
			  $(0.50,-0.50)$ & -0.0078 - 0.0211\ii
			\\ \hline
			    $(-0.50,-0.50)$  & 0.0029 - 0.0215\ii
			\\ \hline
			   $ (0.50,0.50)$  & -0.0078 - 0.0211\ii
			\\ \hline
		      $(-0.50,0.50)$ & 0.0029 - 0.0215\ii \\ 
			\hline
		\end{tabular} 
		\caption{The admittances  $t_\alpha^{-1}$ for the $2\times 2$ array of passive particles sending the wave incident at $\theta = \pi$ for $k a = 5.4$ into the tripole pattern, see Figure \ref{fig33}.  }  \label{table2}
	\end{table}
\end{center}

\begin{figure}[h]
	\centering
	\subcaptionbox{(normalized) Red: pattern at {$ka = 5.3$}, black: target pattern \label{44a} } 
	{\vspace{-.21cm}\includegraphics[width=0.49\linewidth ]{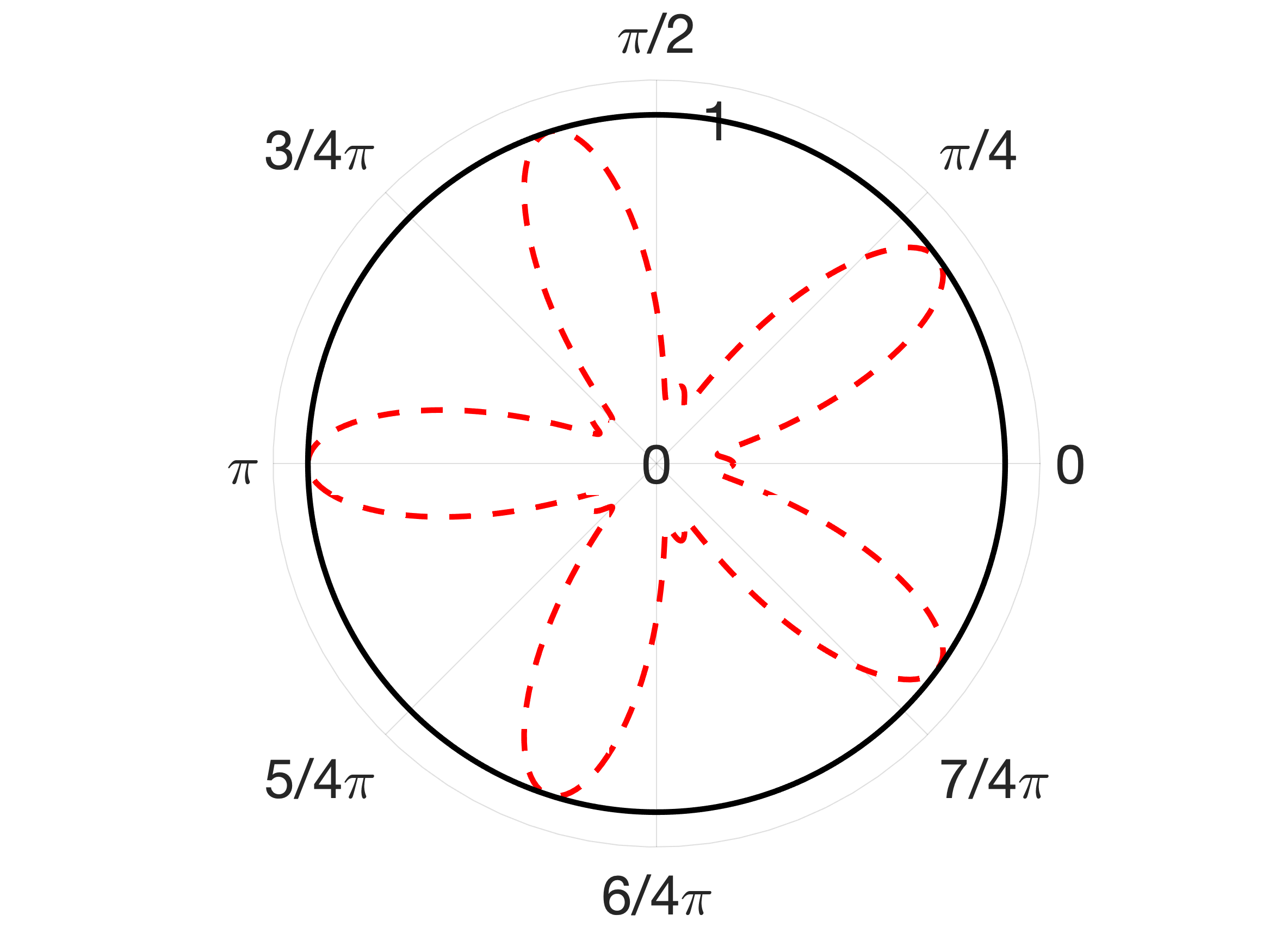}}
	\subcaptionbox{Displacement magnitude at $ka = 5.3$.  } 
	{\vspace{-.21cm}\includegraphics[width=0.49\linewidth  ]{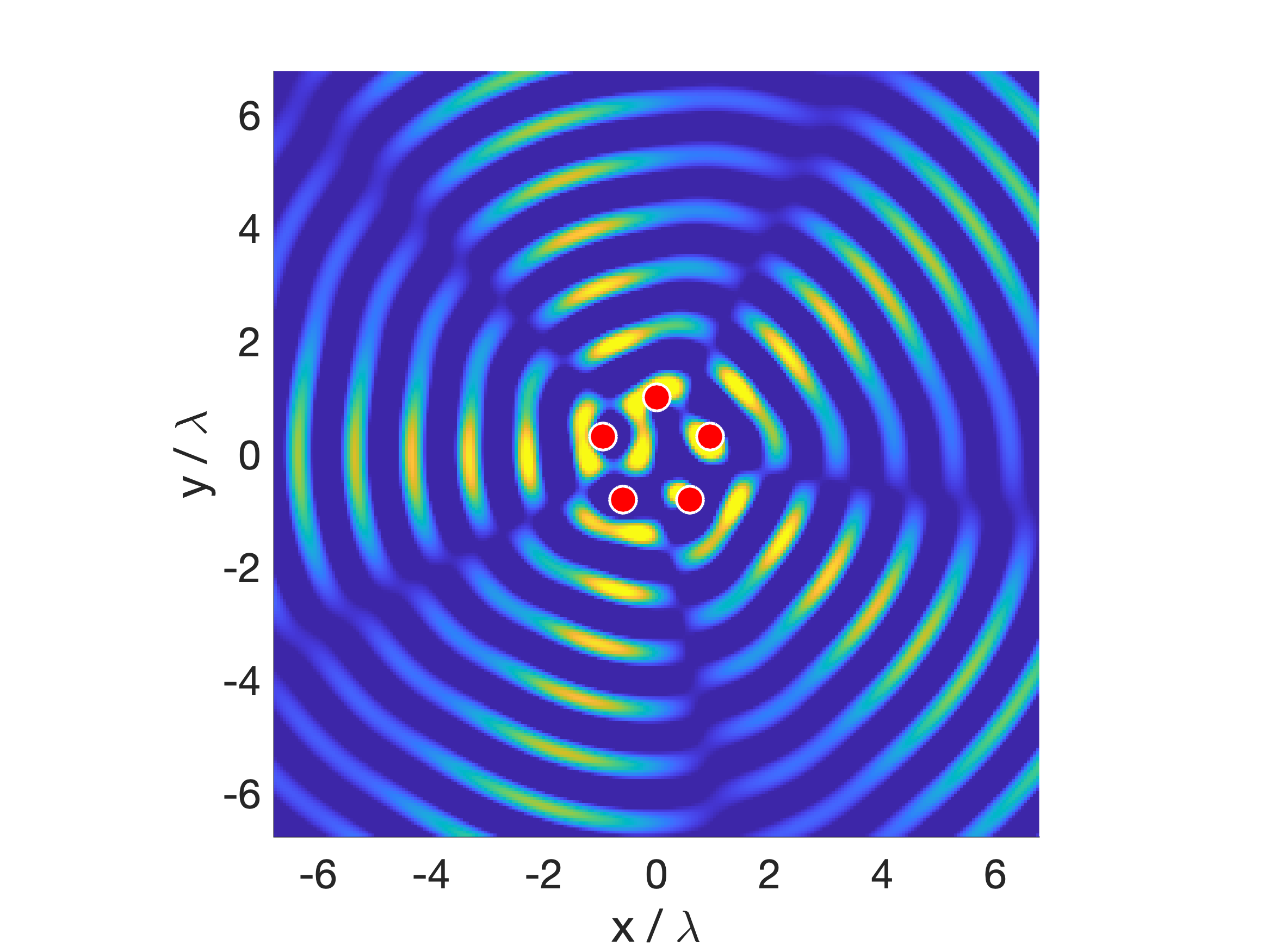}}
	\caption{The optimal pattern function for the passive circular  metacluster of five particles, (a), and the corresponding displacement field generated at $ka = 5.3$, (b). Admittances of the cluster are give in Table \ref{table3}. {$\eta = 0.84$ for this cluster configuration.}}
	\label{fig44}
\end{figure}

Figure \ref{fig44} shows a metacluster designed for generating a pentapole pattern. The cluster consists of a circular arrangement of five scatterers with optimal positions and impedances listed in Table \ref{table3}. Clearly, the cluster is locally passive. It is important to note that this metacluster setup, resulting in nearly perfect pentapole {(red dashed line in Fig. \ref{44a})}, has been obtained accidentally when looking for {the vortex-type scattering pattern} (different than the pentapole pattern, {see black solid line in Fig. \ref{44a}}). The latter is a consequence of relaxing the requirement of enforcing the target phase of the scattered field and indicates that much more complex scattering patterns that are still locally passive may be obtained for desired amplitude-only rather than amplitude-and-phase target fields. {This cluster  also displays high energy conversion efficiency with $\eta = 0.84$.}
   
\begin{center}
	\begin{table}
		\begin{tabular}{| c | c |}
			\hline
			\hline
			${\bf R}_\alpha$ & $t_\alpha^{-1} $ \\
			& $ka = 5.3$ \\
			\hline
			$(0.95, 0.31)$ & 0.0007 - 0.0020\ii
			\\ \hline
			$(0.00, 1.00)$  & 0.0009 - 0.0021\ii
			\\ \hline
			$(-0.95, 0.31)$  & 0.0011 - 0.0020\ii
			\\ \hline
			$(-0.59, -0.81)$ & -0.0045 - 0.0002\ii 
			\\ \hline
			$(0.59, - 0.81)$ & 0.0064 - 0.0003\ii 
			\\ \hline
		\end{tabular} 
		\caption{The admittances  $t_\alpha^{-1}$ for the five-element circular array of passive particles sending the wave incident at $\theta = \pi$ for $k a = 5.3$ into the pentapole pattern, see Figure \ref{fig44}.  }  \label{table3}
	\end{table}
\end{center}

\subsubsection{Example: A passive optimal metacluster for a vortex pattern}

\begin{figure}[h]
	\centering
	\subcaptionbox{Magnitude (normalized). Red: pattern at $ka = 5.5$, black: target pattern \label{22a} } 
	{\vspace{-.21cm}\includegraphics[width=0.49\linewidth ]{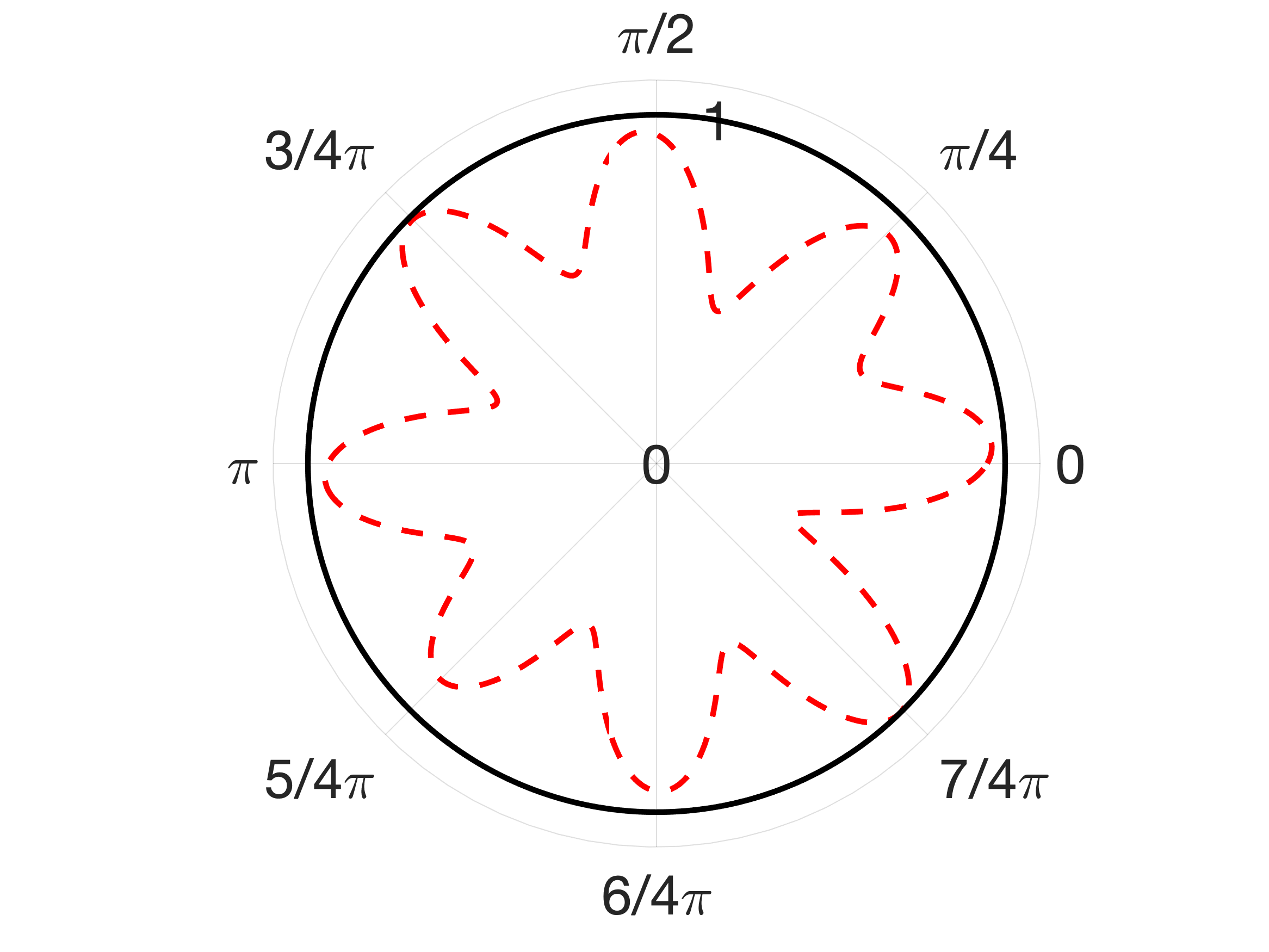}}
	\subcaptionbox{Phase (normalized). Red: pattern at $ka = 5.5$, black: target pattern \label{22b} } 
	{\vspace{-.21cm}\includegraphics[width=0.49\linewidth  ]{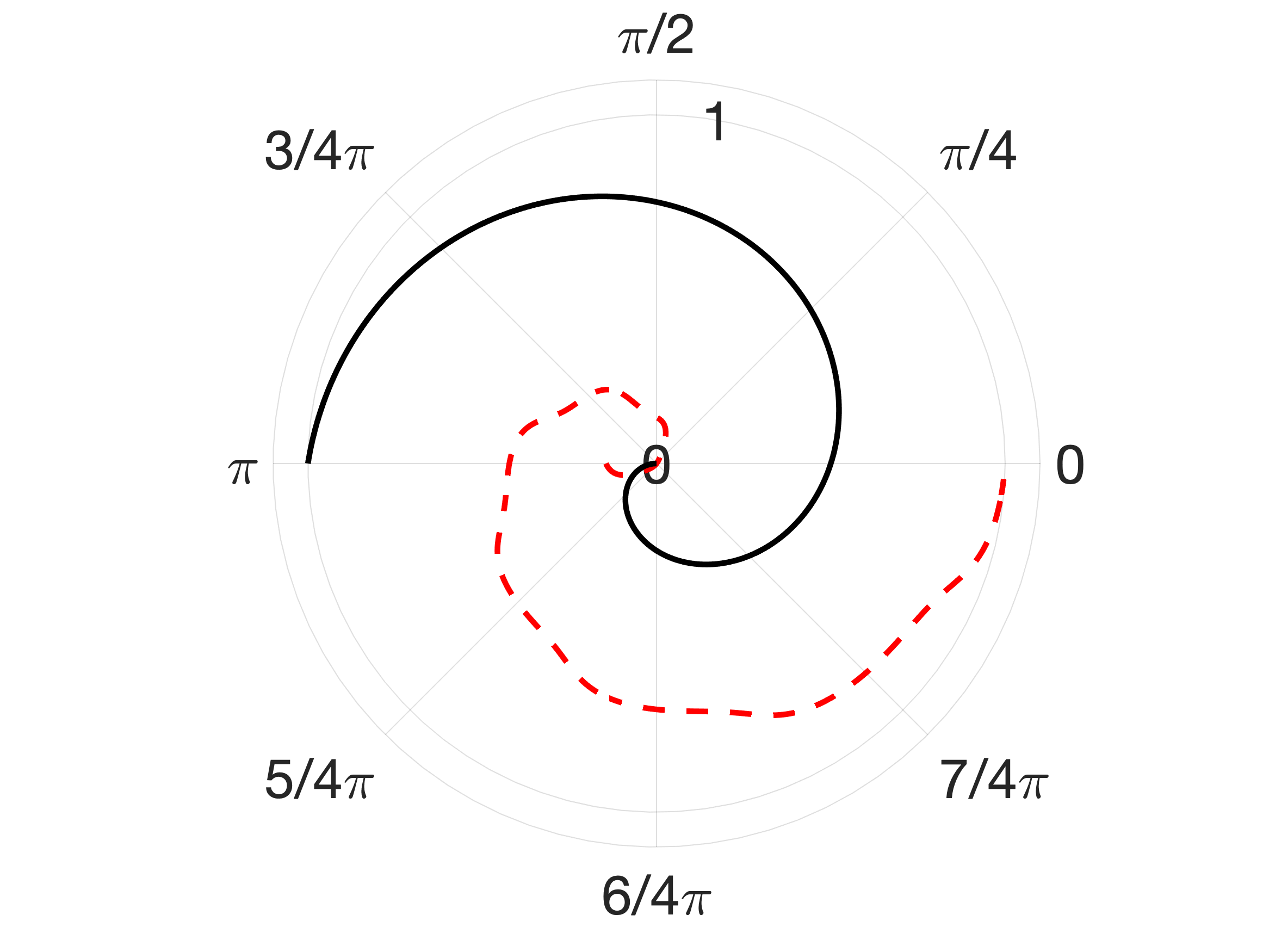}}
	\subcaptionbox{Displacement magnitude at $ka = 5.5$. \label{22c}} 
	{\vspace{-.21cm}\includegraphics[width=0.49\linewidth] {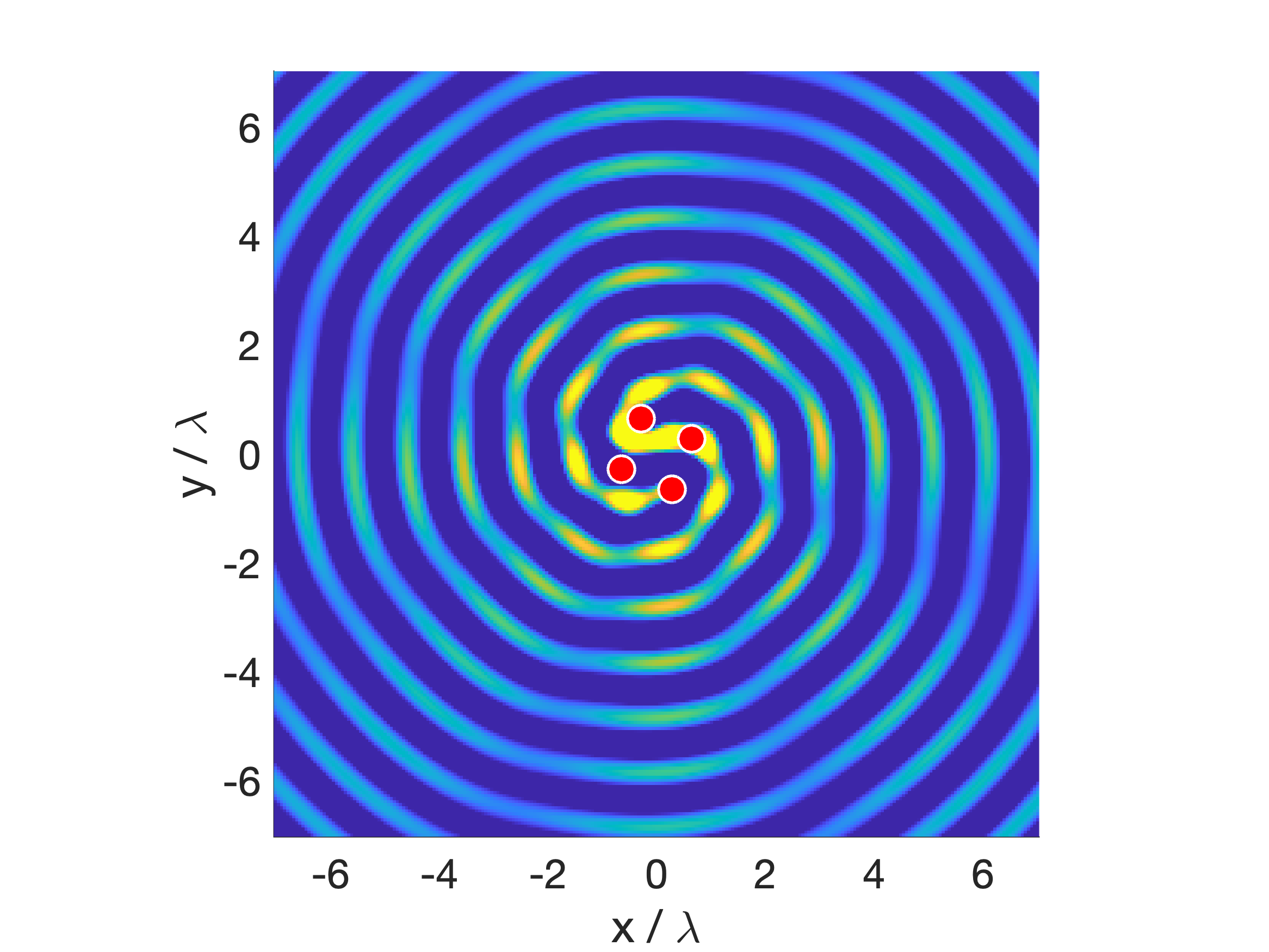}}
	\subcaptionbox{Displacement phase at $ka = 5.5$. \label{22d} } 
	{\vspace{-.21cm}\includegraphics[width=0.49\linewidth]{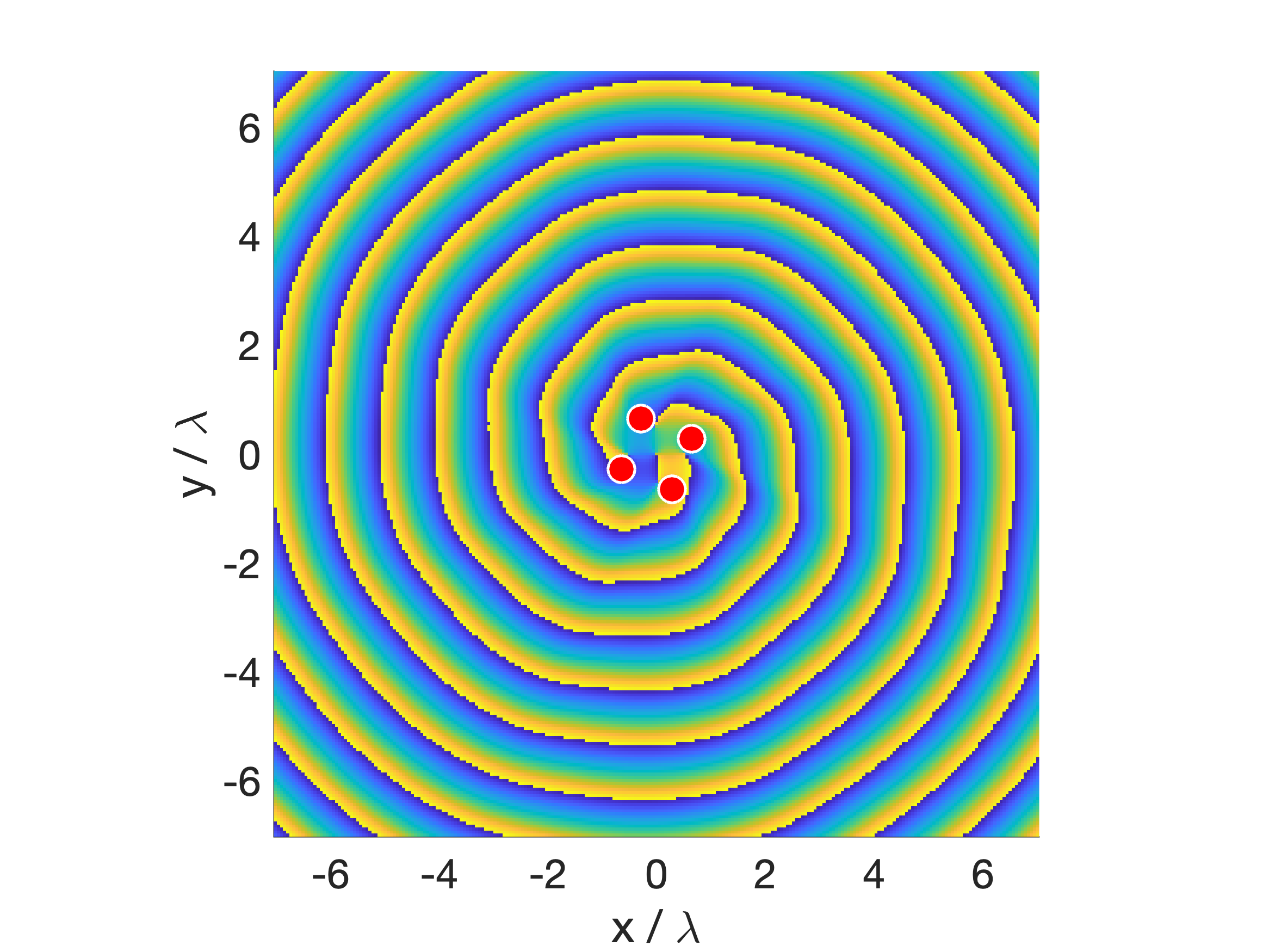}}
	\caption{The optimal pattern function for the passive $2\times2$  metacluster: amplitude - (a) and phase - (b), and the corresponding displacement field generated at $ka = 5.5$: displacement magnitude - (c) and phase - (d). Admittances of the cluster are give in Table  \ref{table4}. {The energy efficiency is $\eta = 0.14$.}}
	\label{fig22}
\end{figure}

Finally, we present a locally passive metacluster capable of transforming the incident wavefield into the first-order vortex, $\bar{p} = 1$, as shown in Fig. \ref{fig22}. It can be seen from Fig. \ref{22a} that despite the amplitude pattern is not perfectly preserved, the phase behavior (Fig. \ref{22b}) recovers the linearly-dependent angular characteristic of the vortex. Figures \ref{22c} and \ref{22d} show displacements and phases of the wavefields generated by the metacluster. {It is worth noting that this relatively complex scattering pattern is obtained by only four passive impedances. The cluster efficiency is $\eta = 0.14$ for this setup, being a consequence of moderate damping in all scatterers.}

\begin{center}
	\begin{table}
		\begin{tabular}{| c | c |}
			\hline
			\hline
			${\bf R}_\alpha$ & $t_\alpha^{-1} $ \\
			& $ka = 5.5$ \\
			\hline
			$(0.28, - 0.65)$ & 0.0683 - 0.0160\ii
			\\ \hline
			$( -0.65, - 0.28 )$  & -0.0658 - 0.0068\ii
			\\ \hline
			$(0.65 ,0.28)$  & 0.0536 - 0.0341\ii
			\\ \hline
			$(  -0.28, 0.65)$ & 0.0676 - 0.01952\ii 
			\\ \hline
		\end{tabular} 
		\caption{The admittances  $t_\alpha^{-1}$ for the $2 \times 2$ square array of passive particles sending the wave incident at $\theta = \pi$ for $k a = 5.5$ into the vortex pattern, see Figure \ref{fig22}.  } \label{table4}
	\end{table}
\end{center}

\section{Summary}
\label{sec:summ} 
We have shown that an inverse multiple scattering method can be applied for the design of the radiation patterns of clusters of scatterers. While the design process is complex and passive solutions are not easy to find, the approach has still more degrees of freedom to explore. The specific case of flexural waves in thin elastic plates has been considered, with the target problem of designing the far field patterns, although it is easy to show that near field patterns can also be considered. Similarly, the multiple scattering formulation is not unique to flexural waves, and the approach introduced here can be easily extended to other classical waves, like optical or acoustical. {Further analysis using clusters of  finite-size scatterers is important for physical realization  of the directivity effect. While the analysis of such attachments requires introduction of  scattering matrices for each object, the structure of the framework proposed here remains unchanged but becomes more involved. However, some simplifications can be made for low frequency approximations of finite-size scatterers, reducing the infinite scattering matrix to only several terms describing monopoles, dipoles, etc. These issues are under current investigation, and we expect to report on the acoustic analog in the near future.}

The proposed metaclusters can be considered a generalization of the notion of a metagrating, where the inverse design is performed in the amplitude of the diffraction orders and the structures are  periodic gratings. {However, in contrast with the infinite number of scattering elements in a metagrating, the present results are based on clusters of very few scatterers.  In light of the small number of  elements employed, the scattering directivity is remarkable in our opinion.}  With the alternative presented here we could design not only finite gratings but also flat lenses, beam splitters and even cloaking devices. We consider therefore that this work opens a new direction towards the design of passive devices for the control of mechanic and electromagnetic energy.

\begin{acknowledgments}

ANN acknowledges the support of the National Science Foundation through EFRI Award no.
1641078.  PP acknowledges the support of the National Science Centre in Poland through grant no. 2018/31/B/ST8/00753.  DT acknowledges financial support through the ``Ram\'on y Cajal'' fellowship under grant number RYC-2016-21188 and to the Ministry of Science, Innovation and Universities through Project No. RTI2018- 093921-A-C42.

\end{acknowledgments}

\appendix 
\section{Plate equations and energy balance}   \label{A}

The plate has thickness $h$, bending stiffness $D$ $(=EI/(1-\nu^2))$ and density $\rho $. 
Time harmonic motion $e^{-\ii \omega t}$ is assumed, so that the flexural wavenumber $k$ is defined by $k^4 = \omega^2 \rho h/D$. 
We assume there are $N$ point scatterers located at 
$ {\bm R}_\alpha  $ 
with impedances $t_\alpha$, $\alpha = 1, 2 , \ldots, N$.  The total displacement $\psi$  satisfies 
\beq{-214}
D\big( \Delta ^2 \psi ({\bf r}) - k^4 \psi ({\bf r}) \big) =  \sum_{\alpha =1}^N 
t_\alpha
\psi ({\bf R}_\alpha) 
\delta({\bf r}-{\bf R}_\alpha)  .
\eeq
A generic  attachment impedance $t$ may be  modeled as single degree of freedom with  mass $M$, spring stiffness  $\kappa$  and damping coefficient $\nu$.  Two possible configurations are 
\beq{-14}
t = 
\begin{cases} 
	\big( \frac 1{M\omega^2} - \frac 1{\kappa - \ii \omega \nu} \big)^{-1},
	& (a) ,
	\\
	M\omega^2 - \kappa + \ii \omega \nu , & (b)  .
\end{cases}
\eeq
In case (a) the mass is attached to the plate by a spring and damper in parallel.  
Model (b) assumes the mass is rigidly attached to the plate, and both are attached to a rigid foundation by the spring and damper in parallel \cite{Evans2007}.  An important limit is a  plate pinned at ${\bm R}_\alpha$, $\psi ({\bm R}_\alpha)=0$, which corresponds to $ t \to \infty$. The (a) and (b) oscillators could also be attached in parallel. e.g.\ on either side of the plate, to give $t= t_a +t_b$. 

The  Green's function is the particular solution $\psi = G$ for a single source of the form $\delta({\bf r})$ on the right hand side of \eqref{-214}, 
\beq{66}
G({\bf r}) = \frac{\ii}{8k^2 D} \Big(
H_0^{(1)}(kr) - H_0^{(1)}(\ii kr) 
\Big) .
\eeq

The following identity may be found starting from the plate equation \eqref{-214} using the procedure of Norris and Vemula \cite{norris1995scattering} for the analogous case without source terms, 
\bal{6=4}
&\Im D
\int_{\partial A} \big( \psi^* ({\bf r}) \nabla\Delta  \psi ({\bf r}) - \Delta  \psi ({\bf r}) \nabla \psi^* ({\bf r}) \big)\cdot {\bf n}\dd s 
\notag \\ 
 \ \ & = \sum_{\alpha =1}^N 
( \Im t_\alpha )
|\psi ({\bf R}_\alpha) |^2  .
\eal
Taking the limit as the bounding surface ${\partial A}$ tends to infinity, and using 
Eqs.\ \eqref{101}, \eqref{-21} and \eqref{-69} yields 
\beq{2--}
\Im f(0) = \frac {\int_0^{2\pi} |f(\theta)|^2 \dd \theta}{16\pi D k^2}
 + \sum_{\alpha =1}^N 
(\Im t_\alpha )
|\psi ({\bf R}_\alpha) |^2  .
\eeq
Define
\bal{3=3}
\sigma_\text{ext} &= \Im f(0) , \notag
\\
\sigma_\text{sca} &=  \frac 1{16\pi D k^2}  {\int_0^{2\pi} |f(\theta)|^2 \dd \theta}, 
\\
\sigma_\text{abs} &=   \sum_{\alpha =1}^N 
(\Im t_\alpha ) |\psi ({\bf R}_\alpha) |^2 , \notag
\eal
then the energy balance   becomes 
\beq{2--2}
\sigma_\text{ext}  = \sigma_\text{sca} +\sigma_\text{abs} . 
\eeq

Note that 
\beq{3=4}
\ba
\sigma_\text{sca} &=  \frac 1{8 D k^2}  {{\bf A}^{(P)}}^\dagger {\bf A}^{(P)} , 
\\
\sigma_\text{abs} &=   \sum_{\alpha =1}^N 
(-\Im t_\alpha^{-1} ) |B_\alpha^{(P)} |^2 , 
\ea
\eeq
where the infinite vector ${\bf A}^{(P)}$ and $N-$vector ${\bf B}^{(P)}$ are the solutions for the passive set of impedances. 

A sort of equivalent reasoning can be derived from Eq.\ \eqref{105} by rewriting it in the matrix form
\beq{105a}
\sum_{\alpha=1}^N t^{-1}_\alpha |B_\alpha|^2 = {\bf B}^\dagger \bm{\psi}_0 + {\bf B}^\dagger  {\bf G}  {\bf B} ,
\eeq
\noindent
where ${\bf G} = \{ G \left( {\bf R}_\alpha - {\bf R}_\beta \right) \}$ and $ \bm{\psi}_0 = \{ \psi_0 \left( \bf{R}_\alpha \right) \}$. From \eqref{105a} we are interested only in the imaginary part, as it defines the passive or active character of the cluster. Note that the imaginary part of the quadratic form in \eqref{105a} is ${\bf B}^\dagger \Im \left( {\bf G} \right) {\bf B}$ and $\Im \left( {\bf G} \right) \propto J_0$, hence $\Im \left( {\bf G} \right)$ is real-valued and symmetric. Finally, for a globally passive cluster we require
\bal{105b}
\sum_{\alpha=1}^N \left( \Im t^{-1}_\alpha \right) |B_\alpha|^2  &  =
{\bf B}^\dagger  \Im \left( {\bf G} \right)  {\bf B}
+\frac{1}{2} \left( {\bf B}^\dagger {\bm{\psi}}_0 - {\bf B}^T  \bm{\psi}_0^* \right)  
\notag \\ & \leq 0 .
\eal

Satisfying $\Im t_\alpha^{-1} \leq 0$ for all individual particles $\alpha$ corresponds to a locally passive metacluster. From Eq.\ \eqref{-214} and \eqref{25-} it may be noted that $B_\alpha$ is a complex force amplitude that acts on the plate. The passivity of a single scatterer can be seen through the Poynting vector - characteristic of the direction of energy flow. Time-averaged energy flow through the point at which a scatterer is placed is
\beq{73f}
\Phi = - \frac{1}{2} \Re \left( B_\alpha \dot{\psi}^\dagger_\alpha \right) = \frac{1}{2} \omega \psi^\dagger_\alpha \psi_\alpha \Im t_\alpha ,
\eeq
\noindent
where for $\Im  t_\alpha \geq 0$ we have $\Phi \geq 0$, so energy flows from the plate towards the scatterer, i.e.\ the scatterer is passive. $\Phi$ can be seen as the power absorbed by the scatterer.

\section{Some matrix properties}   \label{B} 

It follows from the definition of $\bf S$ in \eqref{-4} and Graf's addition theorem for Bessel functions, Eq.\ (9.1.79) of Abramowitz and Stegun \cite{Abramowitz74},  that 
${\bf S}^\dagger {\bf S}$ simplifies to 
\bal{4=5}
\big({\bf S}^\dagger {\bf S}\big)_{\alpha \beta} 
&=  \sum_{n=-\infty}^\infty J_n(kR_\alpha) J_n(kR_\beta) e^{\ii n(\theta_\alpha - \theta_\beta)}
\notag \\
 &=    J_0(kR_{\alpha\beta})  
\eal  
where $R_{\alpha\beta}=|\bm R_\alpha-\bm R_\beta|$.   Note that  $J_0(kR_{\alpha\beta})\approx 1$ at low frequency, indicating that   ${\bf S}^\dagger {\bf S}$ becomes singular in this limit.  Numerical examples  shows this in terms of the matrix condition number which becomes large at low frequency.

The $N\times N$ matrix ${\bf S}^\dagger {\bf S}$ is therefore real, symmetric and non-negative definite, and can be expressed
\beq{4=51}
{\bf S}^\dagger {\bf S} 
=  \sum_{\alpha = 1}^N  \lambda_\alpha {\bf u}_\alpha {\bf u}_\alpha^\dagger 
\eeq  
with positive eigenvalues $\lambda_\alpha>0$ and normalized eigenvectors of length $N$,  ${\bf u}_\alpha^\dagger {\bf u}_\beta = \delta_{\alpha \beta}$. 
Using \eqref{4=51} in the definition of $\bf P$, Eq.\ \eqref{8.1}, yields
\beq{4=52}
{\bf P} 
=  \sum_{\alpha = 1}^N  \lambda_\alpha^{-1} {\bf V}_\alpha {\bf V}_\alpha^\dagger 
\eeq 
where the infinite dimensional vectors ${\bf V}_\alpha$ are 
\beq{4=53} 
 {\bf V}_\alpha = {\bf S}{\bf u}_\alpha , 
\ \ \alpha = 1, \ldots , N.
\eeq 
These are orthogonal, 
$
{\bf V}_\alpha^\dagger {\bf V}_\beta = \lambda_\alpha \delta_{\alpha \beta}$,
but not orthonormal. We  define the orthonormal set
\beq{4=54} 
 {\bf U}_\alpha = \lambda_\alpha^{-1/2} {\bf S}{\bf u}_\alpha , 
\ \ \alpha = 1, \ldots , N, 
\eeq 
so that $\bf P$ is in canonical form, 
\beq{4=55}
{\bf P} 
=  \sum_{\alpha = 1}^N    {\bf U}_\alpha {\bf U}_\alpha^\dagger . 
\eeq 
Hence  ${\bf P} $ is finite 
rank  with $N$ non-zero eigenvalues equal to $+1$.   Alternatively, ${\bf P} $ is a projection onto the $N-$dimensional subspace $\spn \{ {\bf U}_\alpha , 
\, \alpha = 1, \ldots , N\}$, and satisfies the projector property
\beq{4=58}
{\bf P}^2 
=    {\bf P}  . 
\eeq 

We note some other properties of ${\bf P} $ and related matrices. 
 Multiplying \eqref{8.1} on the right by ${\bf S}$  and on the left by ${\bf S}^\dagger$ gives
\beq{8.1a}
{\bf P}{\bf S} = {\bf S}, 
\ \ 
{\bf S}^\dagger {\bf P} = {\bf S}^\dagger .
\eeq
The fundamental matrix $\bf S$ of \eqref{-4} has an interesting form in terms of the finite and infinite dimensional normalized eigenvectors: 
\beq{4-5}
{\bf S} =  \sum_{\alpha = 1}^N   \lambda_\alpha^{1/2}  {\bf U}_\alpha {\bf u}_\alpha^\dagger . 
\eeq
The Moore-Penrose inverse of ${\bf S}$ is 
\beq{6-}
\big( {\bf S}^\dagger {\bf S}\big)^{-1}  {\bf S}^\dagger 
=\sum_{\alpha = 1}^N   \lambda_\alpha^{-1/2}  {\bf u}_\alpha {\bf U}_\alpha^\dagger . 
\eeq

Similarly, the matrix $\bf Q$ of \eqref{12} is 
\beq{4-6}
{\bf Q} =  \sum_{\alpha = 1}^N    {\bf u}_\alpha {\bf U}_\alpha^\dagger . 
\eeq
It follows from this, or from its definition in \eqref{12}, that 
the  matrix $\bf Q$ satisfies  
\beq{13}{\bf Q}^\dagger {\bf Q} = {\bf P}, 
\ \ 
{\bf Q} {\bf Q}^\dagger= {\bf I}_{N}  
\eeq
where  ${\bf I}_{N}  $ is the identity on $\spn \{ {\bf U}_\alpha , \, \alpha = 1, \ldots , N\}$

Finally, the physical vectors for the far field pattern function and source strengths of Eqs.\ \eqref{8} and  \eqref{6} are, respectively, 
\beq{4-61}
\ba
{\bf A}^{(N)} &=  \sum_{\alpha = 1}^N    a_\alpha {\bf U}_\alpha, 
\\
{\bf B} &=  \sum_{\alpha = 1}^N    b_\alpha {\bf u}_\alpha, 
\ea 
\eeq
where 
\beq{4-62}
   a_\alpha  = {\bf U}_\alpha^\dagger {\bf A} , 
\ \
  b_\alpha =\lambda_\alpha^{-1/2} a_\alpha .
\eeq


\begin{thebibliography}{20}%
\makeatletter
\providecommand \@ifxundefined [1]{%
 \@ifx{#1\undefined}
}%
\providecommand \@ifnum [1]{%
 \ifnum #1\expandafter \@firstoftwo
 \else \expandafter \@secondoftwo
 \fi
}%
\providecommand \@ifx [1]{%
 \ifx #1\expandafter \@firstoftwo
 \else \expandafter \@secondoftwo
 \fi
}%
\providecommand \natexlab [1]{#1}%
\providecommand \enquote  [1]{``#1''}%
\providecommand \bibnamefont  [1]{#1}%
\providecommand \bibfnamefont [1]{#1}%
\providecommand \citenamefont [1]{#1}%
\providecommand \href@noop [0]{\@secondoftwo}%
\providecommand \href [0]{\begingroup \@sanitize@url \@href}%
\providecommand \@href[1]{\@@startlink{#1}\@@href}%
\providecommand \@@href[1]{\endgroup#1\@@endlink}%
\providecommand \@sanitize@url [0]{\catcode `\\12\catcode `\$12\catcode
  `\&12\catcode `\#12\catcode `\^12\catcode `\_12\catcode `\%12\relax}%
\providecommand \@@startlink[1]{}%
\providecommand \@@endlink[0]{}%
\providecommand \url  [0]{\begingroup\@sanitize@url \@url }%
\providecommand \@url [1]{\endgroup\@href {#1}{\urlprefix }}%
\providecommand \urlprefix  [0]{URL }%
\providecommand \Eprint [0]{\href }%
\providecommand \doibase [0]{http://dx.doi.org/}%
\providecommand \selectlanguage [0]{\@gobble}%
\providecommand \bibinfo  [0]{\@secondoftwo}%
\providecommand \bibfield  [0]{\@secondoftwo}%
\providecommand \translation [1]{[#1]}%
\providecommand \BibitemOpen [0]{}%
\providecommand \bibitemStop [0]{}%
\providecommand \bibitemNoStop [0]{.\EOS\space}%
\providecommand \EOS [0]{\spacefactor3000\relax}%
\providecommand \BibitemShut  [1]{\csname bibitem#1\endcsname}%
\let\auto@bib@innerbib\@empty
\bibitem [{\citenamefont {Jin}\ \emph {et~al.}(2019{\natexlab{a}})\citenamefont
  {Jin}, \citenamefont {Djafari-Rouhani},\ and\ \citenamefont
  {Torrent}}]{jin2019gradient}%
  \BibitemOpen
  \bibfield  {author} {\bibinfo {author} {\bibfnamefont {Y.}~\bibnamefont
  {Jin}}, \bibinfo {author} {\bibfnamefont {B.}~\bibnamefont
  {Djafari-Rouhani}}, \ and\ \bibinfo {author} {\bibfnamefont {D.}~\bibnamefont
  {Torrent}},\ }\href@noop {} {\bibfield  {journal} {\bibinfo  {journal}
  {Nanophotonics}\ }\textbf {\bibinfo {volume} {8}},\ \bibinfo {pages} {685}
  (\bibinfo {year} {2019}{\natexlab{a}})}\BibitemShut {NoStop}%
\bibitem [{\citenamefont {Engheta}\ and\ \citenamefont
  {Ziolkowski}(2006)}]{engheta2006metamaterials}%
  \BibitemOpen
  \bibfield  {author} {\bibinfo {author} {\bibfnamefont {N.}~\bibnamefont
  {Engheta}}\ and\ \bibinfo {author} {\bibfnamefont {R.~W.}\ \bibnamefont
  {Ziolkowski}},\ }\href@noop {} {\emph {\bibinfo {title} {Metamaterials:
  physics and engineering explorations}}}\ (\bibinfo  {publisher} {John Wiley
  \& Sons},\ \bibinfo {year} {2006})\BibitemShut {NoStop}%
\bibitem [{\citenamefont {Deymier}(2013)}]{deymier2013acoustic}%
  \BibitemOpen
  \bibfield  {author} {\bibinfo {author} {\bibfnamefont {P.~A.}\ \bibnamefont
  {Deymier}},\ }\href@noop {} {\emph {\bibinfo {title} {Acoustic metamaterials
  and phononic crystals}}},\ Vol.\ \bibinfo {volume} {173}\ (\bibinfo
  {publisher} {Springer Science \& Business Media},\ \bibinfo {year}
  {2013})\BibitemShut {NoStop}%
\bibitem [{\citenamefont {Yu}\ \emph {et~al.}(2011)\citenamefont {Yu},
  \citenamefont {Genevet}, \citenamefont {Kats}, \citenamefont {Aieta},
  \citenamefont {Tetienne}, \citenamefont {Capasso},\ and\ \citenamefont
  {Gaburro}}]{yu2011light}%
  \BibitemOpen
  \bibfield  {author} {\bibinfo {author} {\bibfnamefont {N.}~\bibnamefont
  {Yu}}, \bibinfo {author} {\bibfnamefont {P.}~\bibnamefont {Genevet}},
  \bibinfo {author} {\bibfnamefont {M.~A.}\ \bibnamefont {Kats}}, \bibinfo
  {author} {\bibfnamefont {F.}~\bibnamefont {Aieta}}, \bibinfo {author}
  {\bibfnamefont {J.-P.}\ \bibnamefont {Tetienne}}, \bibinfo {author}
  {\bibfnamefont {F.}~\bibnamefont {Capasso}}, \ and\ \bibinfo {author}
  {\bibfnamefont {Z.}~\bibnamefont {Gaburro}},\ }\href@noop {} {\bibfield
  {journal} {\bibinfo  {journal} {science}\ }\textbf {\bibinfo {volume}
  {334}},\ \bibinfo {pages} {333} (\bibinfo {year} {2011})}\BibitemShut
  {NoStop}%
\bibitem [{\citenamefont {Kildishev}\ \emph {et~al.}(2013)\citenamefont
  {Kildishev}, \citenamefont {Boltasseva},\ and\ \citenamefont
  {Shalaev}}]{kildishev2013planar}%
  \BibitemOpen
  \bibfield  {author} {\bibinfo {author} {\bibfnamefont {A.~V.}\ \bibnamefont
  {Kildishev}}, \bibinfo {author} {\bibfnamefont {A.}~\bibnamefont
  {Boltasseva}}, \ and\ \bibinfo {author} {\bibfnamefont {V.~M.}\ \bibnamefont
  {Shalaev}},\ }\href@noop {} {\bibfield  {journal} {\bibinfo  {journal}
  {Science}\ }\textbf {\bibinfo {volume} {339}} (\bibinfo {year}
  {2013})}\BibitemShut {NoStop}%
\bibitem [{\citenamefont {Yu}\ and\ \citenamefont
  {Capasso}(2014)}]{yu2014flat}%
  \BibitemOpen
  \bibfield  {author} {\bibinfo {author} {\bibfnamefont {N.}~\bibnamefont
  {Yu}}\ and\ \bibinfo {author} {\bibfnamefont {F.}~\bibnamefont {Capasso}},\
  }\href@noop {} {\bibfield  {journal} {\bibinfo  {journal} {Nature materials}\
  }\textbf {\bibinfo {volume} {13}},\ \bibinfo {pages} {139} (\bibinfo {year}
  {2014})}\BibitemShut {NoStop}%
\bibitem [{\citenamefont {Ra'di}\ \emph {et~al.}(2017)\citenamefont {Ra'di},
  \citenamefont {Sounas},\ and\ \citenamefont {Al{\`u}}}]{ra2017metagratings}%
  \BibitemOpen
  \bibfield  {author} {\bibinfo {author} {\bibfnamefont {Y.}~\bibnamefont
  {Ra'di}}, \bibinfo {author} {\bibfnamefont {D.~L.}\ \bibnamefont {Sounas}}, \
  and\ \bibinfo {author} {\bibfnamefont {A.}~\bibnamefont {Al{\`u}}},\
  }\href@noop {} {\bibfield  {journal} {\bibinfo  {journal} {Physical review
  letters}\ }\textbf {\bibinfo {volume} {119}},\ \bibinfo {pages} {067404}
  (\bibinfo {year} {2017})}\BibitemShut {NoStop}%
\bibitem [{\citenamefont {Wong}\ and\ \citenamefont
  {Eleftheriades}(2018)}]{wong2018perfect}%
  \BibitemOpen
  \bibfield  {author} {\bibinfo {author} {\bibfnamefont {A.~M.}\ \bibnamefont
  {Wong}}\ and\ \bibinfo {author} {\bibfnamefont {G.~V.}\ \bibnamefont
  {Eleftheriades}},\ }\href@noop {} {\bibfield  {journal} {\bibinfo  {journal}
  {Physical Review X}\ }\textbf {\bibinfo {volume} {8}},\ \bibinfo {pages}
  {011036} (\bibinfo {year} {2018})}\BibitemShut {NoStop}%
\bibitem [{\citenamefont {Torrent}(2018)}]{torrent2018acoustic}%
  \BibitemOpen
  \bibfield  {author} {\bibinfo {author} {\bibfnamefont {D.}~\bibnamefont
  {Torrent}},\ }\href@noop {} {\bibfield  {journal} {\bibinfo  {journal}
  {Physical Review B}\ }\textbf {\bibinfo {volume} {98}},\ \bibinfo {pages}
  {060101} (\bibinfo {year} {2018})}\BibitemShut {NoStop}%
\bibitem [{\citenamefont {Packo}\ \emph {et~al.}(2019)\citenamefont {Packo},
  \citenamefont {Norris},\ and\ \citenamefont {Torrent}}]{packo2019inverse}%
  \BibitemOpen
  \bibfield  {author} {\bibinfo {author} {\bibfnamefont {P.}~\bibnamefont
  {Packo}}, \bibinfo {author} {\bibfnamefont {A.~N.}\ \bibnamefont {Norris}}, \
  and\ \bibinfo {author} {\bibfnamefont {D.}~\bibnamefont {Torrent}},\
  }\href@noop {} {\bibfield  {journal} {\bibinfo  {journal} {Physical Review
  Applied}\ }\textbf {\bibinfo {volume} {11}},\ \bibinfo {pages} {014023}
  (\bibinfo {year} {2019})}\BibitemShut {NoStop}%
\bibitem [{\citenamefont {Popov}\ \emph
  {et~al.}(2019{\natexlab{a}})\citenamefont {Popov}, \citenamefont {Boust},\
  and\ \citenamefont {Burokur}}]{popov2019constructing}%
  \BibitemOpen
  \bibfield  {author} {\bibinfo {author} {\bibfnamefont {V.}~\bibnamefont
  {Popov}}, \bibinfo {author} {\bibfnamefont {F.}~\bibnamefont {Boust}}, \ and\
  \bibinfo {author} {\bibfnamefont {S.~N.}\ \bibnamefont {Burokur}},\
  }\href@noop {} {\bibfield  {journal} {\bibinfo  {journal} {Physical Review
  Applied}\ }\textbf {\bibinfo {volume} {11}},\ \bibinfo {pages} {024074}
  (\bibinfo {year} {2019}{\natexlab{a}})}\BibitemShut {NoStop}%
\bibitem [{\citenamefont {Popov}\ \emph
  {et~al.}(2019{\natexlab{b}})\citenamefont {Popov}, \citenamefont {Yakovleva},
  \citenamefont {Boust}, \citenamefont {Pelouard}, \citenamefont {Pardo},\ and\
  \citenamefont {Burokur}}]{popov2019designing}%
  \BibitemOpen
  \bibfield  {author} {\bibinfo {author} {\bibfnamefont {V.}~\bibnamefont
  {Popov}}, \bibinfo {author} {\bibfnamefont {M.}~\bibnamefont {Yakovleva}},
  \bibinfo {author} {\bibfnamefont {F.}~\bibnamefont {Boust}}, \bibinfo
  {author} {\bibfnamefont {J.-L.}\ \bibnamefont {Pelouard}}, \bibinfo {author}
  {\bibfnamefont {F.}~\bibnamefont {Pardo}}, \ and\ \bibinfo {author}
  {\bibfnamefont {S.~N.}\ \bibnamefont {Burokur}},\ }\href@noop {} {\bibfield
  {journal} {\bibinfo  {journal} {Physical Review Applied}\ }\textbf {\bibinfo
  {volume} {11}},\ \bibinfo {pages} {044054} (\bibinfo {year}
  {2019}{\natexlab{b}})}\BibitemShut {NoStop}%
\bibitem [{\citenamefont {Jin}\ \emph {et~al.}(2019{\natexlab{b}})\citenamefont
  {Jin}, \citenamefont {Fang}, \citenamefont {Li},\ and\ \citenamefont
  {Torrent}}]{jin2019engineered}%
  \BibitemOpen
  \bibfield  {author} {\bibinfo {author} {\bibfnamefont {Y.}~\bibnamefont
  {Jin}}, \bibinfo {author} {\bibfnamefont {X.}~\bibnamefont {Fang}}, \bibinfo
  {author} {\bibfnamefont {Y.}~\bibnamefont {Li}}, \ and\ \bibinfo {author}
  {\bibfnamefont {D.}~\bibnamefont {Torrent}},\ }\href@noop {} {\bibfield
  {journal} {\bibinfo  {journal} {Physical Review Applied}\ }\textbf {\bibinfo
  {volume} {11}},\ \bibinfo {pages} {011004} (\bibinfo {year}
  {2019}{\natexlab{b}})}\BibitemShut {NoStop}%
\bibitem [{\citenamefont {Ni}\ \emph {et~al.}(2019)\citenamefont {Ni},
  \citenamefont {Fang}, \citenamefont {Hou}, \citenamefont {Li},\ and\
  \citenamefont {Assouar}}]{ni2019high}%
  \BibitemOpen
  \bibfield  {author} {\bibinfo {author} {\bibfnamefont {H.}~\bibnamefont
  {Ni}}, \bibinfo {author} {\bibfnamefont {X.}~\bibnamefont {Fang}}, \bibinfo
  {author} {\bibfnamefont {Z.}~\bibnamefont {Hou}}, \bibinfo {author}
  {\bibfnamefont {Y.}~\bibnamefont {Li}}, \ and\ \bibinfo {author}
  {\bibfnamefont {B.}~\bibnamefont {Assouar}},\ }\href@noop {} {\bibfield
  {journal} {\bibinfo  {journal} {Physical Review B}\ }\textbf {\bibinfo
  {volume} {100}},\ \bibinfo {pages} {104104} (\bibinfo {year}
  {2019})}\BibitemShut {NoStop}%
\bibitem [{\citenamefont {He}\ \emph {et~al.}(2020)\citenamefont {He},
  \citenamefont {Jiang}, \citenamefont {Ta},\ and\ \citenamefont
  {Wang}}]{he2020experimental}%
  \BibitemOpen
  \bibfield  {author} {\bibinfo {author} {\bibfnamefont {J.}~\bibnamefont
  {He}}, \bibinfo {author} {\bibfnamefont {X.}~\bibnamefont {Jiang}}, \bibinfo
  {author} {\bibfnamefont {D.}~\bibnamefont {Ta}}, \ and\ \bibinfo {author}
  {\bibfnamefont {W.}~\bibnamefont {Wang}},\ }\href@noop {} {\bibfield
  {journal} {\bibinfo  {journal} {Applied Physics Letters}\ }\textbf {\bibinfo
  {volume} {117}},\ \bibinfo {pages} {091901} (\bibinfo {year}
  {2020})}\BibitemShut {NoStop}%
\bibitem [{\citenamefont {Martin}(2006)}]{martin2006multiple}%
  \BibitemOpen
  \bibfield  {author} {\bibinfo {author} {\bibfnamefont {P.~A.}\ \bibnamefont
  {Martin}},\ }\href@noop {} {\emph {\bibinfo {title} {Multiple scattering:
  interaction of time-harmonic waves with N obstacles}}},\ \bibinfo {number}
  {107}\ (\bibinfo  {publisher} {Cambridge University Press},\ \bibinfo {year}
  {2006})\BibitemShut {NoStop}%
\bibitem [{\citenamefont {Torrent}\ \emph {et~al.}(2013)\citenamefont
  {Torrent}, \citenamefont {Mayou},\ and\ \citenamefont
  {S{\'a}nchez-Dehesa}}]{torrent2013elastic}%
  \BibitemOpen
  \bibfield  {author} {\bibinfo {author} {\bibfnamefont {D.}~\bibnamefont
  {Torrent}}, \bibinfo {author} {\bibfnamefont {D.}~\bibnamefont {Mayou}}, \
  and\ \bibinfo {author} {\bibfnamefont {J.}~\bibnamefont
  {S{\'a}nchez-Dehesa}},\ }\href@noop {} {\bibfield  {journal} {\bibinfo
  {journal} {Physical Review B}\ }\textbf {\bibinfo {volume} {87}},\ \bibinfo
  {pages} {115143} (\bibinfo {year} {2013})}\BibitemShut {NoStop}%
\bibitem [{\citenamefont {Norris}\ and\ \citenamefont
  {Vemula}(1995)}]{norris1995scattering}%
  \BibitemOpen
  \bibfield  {author} {\bibinfo {author} {\bibfnamefont {A.}~\bibnamefont
  {Norris}}\ and\ \bibinfo {author} {\bibfnamefont {C.}~\bibnamefont
  {Vemula}},\ }\href@noop {} {\bibfield  {journal} {\bibinfo  {journal}
  {Journal of sound and vibration}\ }\textbf {\bibinfo {volume} {181}},\
  \bibinfo {pages} {115} (\bibinfo {year} {1995})}\BibitemShut {NoStop}%
\bibitem [{\citenamefont {Evans}\ and\ \citenamefont
  {Porter.}(2007)}]{Evans2007}%
  \BibitemOpen
  \bibfield  {author} {\bibinfo {author} {\bibfnamefont {D.~V.}\ \bibnamefont
  {Evans}}\ and\ \bibinfo {author} {\bibfnamefont {R.}~\bibnamefont
  {Porter.}},\ }\href@noop {} {\bibfield  {journal} {\bibinfo  {journal}
  {Journal of Engineering Mathematics}\ }\textbf {\bibinfo {volume} {58}},\
  \bibinfo {pages} {317} (\bibinfo {year} {2007})}\BibitemShut {NoStop}%
\bibitem [{\citenamefont {Abramowitz}\ and\ \citenamefont
  {Stegun}(1974)}]{Abramowitz74}%
  \BibitemOpen
  \bibfield  {author} {\bibinfo {author} {\bibfnamefont {M.}~\bibnamefont
  {Abramowitz}}\ and\ \bibinfo {author} {\bibfnamefont {I.}~\bibnamefont
  {Stegun}},\ }\href@noop {} {\emph {\bibinfo {title} {Handbook of Mathematical
  Functions with Formulas, Graphs, and Mathematical Tables}}}\ (\bibinfo
  {publisher} {Dover, New York},\ \bibinfo {year} {1974})\BibitemShut {NoStop}%
\end{thebibliography}
%


\end{document}